\documentclass[12pt]{iopart}

\usepackage{iopams}  
\usepackage{graphicx}
\usepackage{dcolumn}
\usepackage{bm}
\usepackage{bbm}
\usepackage{braket}
\usepackage{xcolor}
\usepackage{appendix}

\newcommand{\car}{$^{13}$C~}

\begin{document}

\title{Characterizing the magnetic noise power spectrum of dark spins in diamond}

\author{Ethan Q. Williams}
\ead{ethan.q.williams.gr@dartmouth.edu}
\author{Chandrasekhar Ramanathan}
\ead{chandrasekhar.ramanathan@dartmouth.edu}

\vspace{10pt}

\address{Department of Physics and Astronomy, Dartmouth College, Hanover, New Hampshire 03755, USA}


\vspace{10pt}

\begin{indented}
\item[]August 2024
\end{indented}

\begin{abstract}

Coherence times of spin qubits in solid-state platforms are often limited by the presence of a spin bath. While some properties of these typically dark bath spins can be indirectly characterized via the central qubit, it is important to characterize their properties by direct measurement. Here we use pulsed electron paramagnetic resonance (pEPR) based Carr-Purcell-Meiboom-Gill (CPMG) dynamical decoupling to measure the magnetic noise power spectra for ensembles of P1 (substitutional nitrogen) centers in diamond that typically form the bath for NV (nitrogen-vacancy) centers. The experiments on the P1 centers were performed on a low [N] CVD (chemical vapor deposition) sample and a high [N] HPHT (high-temperature, high-pressure) sample at 89 mT. We characterize the NV centers of the latter sample using the same 2.5 GHz pEPR spectrometer. All power spectra show two distinct features, a broad component that is observed to scale as approximately $1/\omega^{0.7-1.0}$, and a prominent peak at the $^{13}$C Larmor frequency. The behavior of the broad component is consistent with an inhomogeneous distribution of Lorentzian spectra due to clustering of P1 centers, which has recently been shown to be prevalent in HPHT diamond. It is unknown if such clustering occurs in CVD diamond. We develop techniques utilizing harmonics of the CPMG filter function to improve characterization of high-frequency signals, which we demonstrate on the $^{13}$C nuclear Larmor frequency. At 190 mT this is 2.04 MHz, 5.7 times higher than the CPMG modulation frequency ($<357$ kHz, hardware-limited). We assess the robustness of our methods in the presence of finite pulse widths and flip angle errors. Understanding the interactions of dark spins will inform methods of diamond fabrication for quantum technology. These techniques are applicable to ac magnetometry for nanoscale nuclear magnetic resonance and chemical sensing.
\end{abstract}

%
\vspace{2pc}

%
%
%
%

\section{Introduction}
Dynamical decoupling (DD) \cite{carr_effects_1954, meiboom_modified_1958} noise spectroscopy (NS) has become an important tool in characterizing qubit environments in different physical platforms \cite{viola_dynamical_1999, uhrig_keeping_2007, cywinski_how_2008, yuge_measurement_2011, biercuk_dynamical_2011, pham_enhanced_2012, norris_qubit_2016, degen_quantum_2017, ajoy_optimal_2011, alvarez_measuring_2011, harbridge_comparison_2003, mitrikas_extending_2014, mitrikas_modulation_2015, fu_molecular-spin-qubit_2021, bylander_noise_2011, sung_non-gaussian_2019}. For spin-based systems such as nitrogen-vacancy (NV) centers in diamond, optically detected magnetic resonance (ODMR) has been used to characterize the magnetic noise fluctuations around both single centers and ensembles \cite{de_lange_universal_2010, ryan_robust_2010, bar-gill_suppression_2012, wang_spin_2013, staudacher_nuclear_2013, rosskopf_investigation_2014, myers_probing_2014, zhao_dynamical_2014, loretz_spurious_2015, lang_enhanced_2017, romach_measuring_2019, smits_two-dimensional_2019, li_determination_2021, sun_self-consistent_2022, silani_nuclear_2023}. In solid-state spin systems the magnetic fluctuations are often produced by other spins -- both electronic and nuclear -- in the system. For example, it is the interactions with \car nuclear spins and the electronic spins of substitutional nitrogen (P1) centers that dominate the magnetic fluctuations seen by NV centers in bulk diamond.  While the dynamics of these ``dark" spins can be indirectly probed via their interactions with the ``bright" NV center, it is informative to apply DD NS to the dark spins to directly characterize their properties. Furthermore, dark spins in diamond are a promising platform for studying the many-body physics of interacting spin systems \cite{zu_emergent_2021, davis_probing_2023} and may be used as qubits for quantum sensing and computation \cite{degen_entanglement_2021}. The ability to directly probe the properties of dark spins is key to helping understand the coherence properties of new spin qubit platforms based on molecular magnets \cite{gaita-arino_molecular_2019,chilton_molecular_2022} and embedded radical systems \cite{wu_covalent_2018, fataftah_trigonal_2020, jellen_2d_2020}.

For diamond-based quantum sensing and information applications, nitrogen concentrations between approximately 1 ppm and 100 ppm play a non-trivial role in the spin coherence dynamics  \cite{wyk_dependences_1997, bauch_decoherence_2020, barry_sensitivity_2020}. 
The $T_2$ of the NV center has been used to estimate the local P1 concentration 
via ensemble experiments on isotopically engineered CVD (chemical vapor deposition) diamond \cite{bauch_decoherence_2020}.
However, such scaling relationships need to be used carefully as the distribution of the P1 centers is not always uniform.  For example, it has recently been shown that high pressure, high temperature (HPHT) diamond with a reported nitrogen concentration of [N] $\approx$ 200 ppm that was irradiated and annealed for the creation of NV centers had local P1 concentrations varying as widely as 10 to 300 ppm \cite{li_determination_2021}.  Clustering of P1 spins in HPHT diamond has also been observed via pEPR \cite{bussandri_p1_2023, nir-arad_nitrogen_2023} and gives rise to multiple mechanisms for dynamic nuclear polarization \cite{shimon_large_2022}. Characterizing the properties of the P1 centers improves our understanding of the variety of spin environments found in diamond.

\begin{figure*}[ht]
\includegraphics[width=\linewidth]{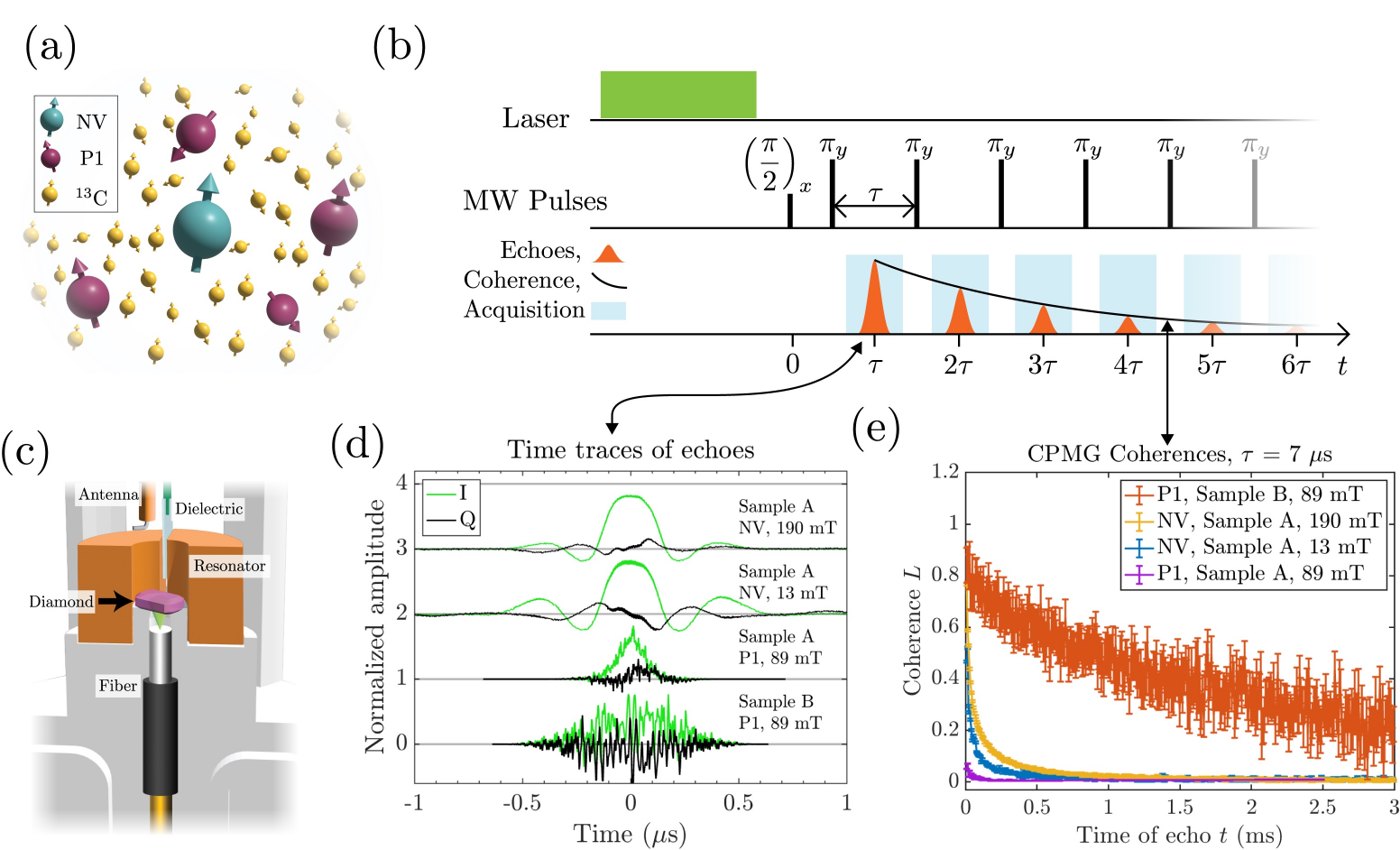}
\caption{\label{fig:1} (a) Cartoon of the central spin model, depicting the NV center as the central spin of a disordered system. (b) Schematic of the CPMG experiment protocol. The laser is used only for NV experiments. The $N^\textrm{th}$ 
echo occurs at detection time $t =N\tau$. In a single shot, we acquire each echo that forms until the signal has diminished below the noise floor. The decay of the echo amplitudes is represented by the coherence curve. (c) Illustration of the assembly of the diamond sample, loop-gap resonator (cutaway view), optical fiber (emitting green 532 nm light), microwave antenna, and sapphire dielectric used for tuning the resonator. (d) Example time traces of spin echo for each of the spin species targeted in this work. For each time trace, the in-phase and quadrature (`I' and `Q') components are shown.  Sample A is the HPHT diamond and Sample B is the CVD diamond. (e) Examples of CPMG coherence decays with fixed $\tau = 7 \ \mu$s.}
\end{figure*}

DD NS of dark spins can be achieved using inductively-detected pulsed electron paramagnetic resonance (pEPR). 
Here we demonstrate the use of pEPR to characterize the noise spectrum of ensembles of P1 centers at these two extremes with a relatively low [N] CVD sample and a relatively high [N] HPHT sample. We show that pEPR can also be used to characterize the noise spectrum of the NV centers in the HPHT sample, allowing us to compare our results to those of previous ODMR measurements.  While typically performed on large ensembles of spins, inductive detection of small ensembles on the order of 1000 spins can also be performed \cite{artzi_induction-detection_2015, blank_recent_2017}.

We extend standard DD NS analysis techniques to reconstruct the magnetic noise power spectra seen by the spins.   All power spectra  showed two distinct features, a broad component that is observed to scale as approximately $1/\omega^{0.7-1.0}$, and a prominent peak at the $^{13}$C Larmor frequency.   We develop techniques that utilize the higher harmonics of the CPMG (Carr-Purcell-Meiboom-Gill \cite{carr_effects_1954, meiboom_modified_1958}) filter function to improve our ability to characterize peaks in the power spectrum at frequencies higher than the CPMG maximum modulation frequency (about 360 kHz with our hardware). 
For the NV experiments at 190 mT, the $^{13}$C Larmor frequency is 5.7 times higher than the CPMG modulation frequency.  We assess the robustness of our methods in the presence of finite pulse widths and flip angle errors.

\section{Experiment Overview}

\label{Sec:ExpOvrw}

The experiments are performed on a 2.5 GHz lab-built pEPR spectrometer.  We can access 3 resonance configurations at this frequency. For the negatively charged spin-1 NV center, the $\ket{0} \leftrightarrow \ket{-1}$ resonance frequency crosses 2.5 GHz at $B_0 = 13$ mT and 190 mT. For the spin-1/2 P1 center the $\ket{-1/2} \leftrightarrow \ket{1/2}$ resonance occurs at 89 mT.  For both P1 and NV centers, hyperfine interactions with the host $^{14}$N nucleus (spin-1), result in three nuclear spin manifolds, and we set our magnetic field such that our pulses are on resonance with the central ($m_I=0$) manifold.

A cartoon of the spin environment in diamond is shown in Figure~\ref{fig:1}(a).  There are two diamond samples used in this work, both purchased from Element 6. Sample A is a high pressure, high temperature (HPHT) Type Ib diamond with a stone-type cut, approximate dimensions $4.2 \times 5.7 \times 1.3$ mm$^3$, and nitrogen and nitrogen-vacancy concentrations measured to be [N] = 87(8) ppm and [NV] = 1.7(3) ppm. Sample B is a chemical vapor deposition (CVD) diamond plate with approximate dimensions $3.25 \times 3.17 \times 0.28$ mm$^3$ and [N] = 0.39(9) ppm. A full description of the samples is given in Supplementary Information S1.

A schematic of the CPMG sequence is shown in Figure \ref{fig:1}(b). In a single CPMG experiment with a fixed pulse-spacing $\tau$, we stroboscopically acquire each echo formed.  Figure \ref{fig:1}(d) shows examples of time traces of single echoes from the four spin species addressed in this work, while Figure~\ref{fig:1}(e) shows example CPMG decay envelopes.

The microwave excitation and detection is performed inductively using a loop-gap resonator \cite{hardy_splitring_1981, froncisz_loop-gap_1982, joshi_adjustable_2020}. We use a capacitive antenna coupling to the resonator which houses the diamond sample as depicted in  Figure~\ref{fig:1}(c). Details of the experimental setup are provided in Supplementary Information S3.

\subsection{P1 Centers}
In the P1 center, the $^{14}$N hyperfine interactions are strong, and the $m_I = +1$ or $-1$ resonance lines are outside the bandwidth of the pEPR pulse.  A Jahn-Teller distortion can generate up to 4 spectral lines for each state $m_I$ of the host nitrogen \cite{smith_electron-spin_1959, bauch_ultralong_2018}. For the P1 experiments on Sample A (B), the sample is oriented with the [111] ([100]) axis aligned with $\vec{B}_0$. For both sample alignments, the 4 spectral lines of the $m_I=0$ manifold spectrally overlap and are indistinguishable as evidenced by echo-detected field sweep spectra shown in Supplementary Information S2.

P1 CPMG data for Sample A (B) was acquired with 16,384 (65,536) repetitions for averaging. High amounts of averaging are required to obtain sufficient signal-to-noise due to the low thermal spin polarization.  At $T=300 \textrm{ K}$ the thermal spin polarization of the P1 centers is  $\epsilon_\textrm{th} \approx 2.0 \times 10^{-4}.$

\subsection{NV Centers}
For the NV experiments, Sample A is oriented so that the [111] axis is aligned with $\vec{B}_0$. Here, the three hyperfine lines are within the bandwidth of our $\sim$50 ns microwave pulse as evidenced by the echoes and spectra in Figures S2 and S3 of Supplementary Information S2. The NV centers in the diamond sample are initialized to the $\ket{m_s}=\ket{0}$ ground state with a 10 ms pulse of 532 nm unfocused light transmitted via an optical fiber. The intensity of the light incident on the diamond is 1 W/mm$^2$ and directly illuminates approximately 1/3 of the sample, while the light that is reflected and diffused by the sample and sample holders illuminates the rest. The mean hyperpolarization with green light, calculated from the observed signal enhancement, is $\epsilon_\textrm{hyp} \approx 120 \epsilon_\textrm{th}$.  As a result only 128 repetitions were needed for the NV center experiments.

\section{Dynamical decoupling-based spectroscopy in diamond}
\label{Sec:DDSpec}
\subsection{System and decoherence model in diamond}
For the NV spins in Sample A and P1 spins in Sample B the concentration of the measured spins is low and the dipolar interactions between identical measured spins can be ignored.  In these cases the underlying physics for both systems can be described in terms of the central spin model \cite{yang_quantum_2016}.  Here the environment consists of a spin bath as we can ignore interactions with lattice phonons on timescales shorter than the spin-lattice relaxation time $T_1$.  Note, however, this model breaks down when considering the P1 spins in Sample A, where dipolar interactions between the P1 spins can no longer be  ignored.

For an electronic central spin in diamond (in the $m_I=0$ $^{14}$N manifold), the total Hamiltonian can be expressed as
\begin{equation}
    H_\textrm{Tot}(t) = H_S + H_C(t) + H_E + H_{SE}
    \label{eqn:TotHam}
\end{equation}
where $H_S$, the system Hamiltonian, determines the effective resonance frequency of the system under study and so contains the Zeeman interaction with the external magnetic field, and, in the case of the NV center, the zero-field splitting. The control Hamiltonian is $H_C(t)=-\gamma_e \vec{S} \cdot \vec{B}_1(t)$, where $\vec{S}$ is the central spin operator vector $(S_x,S_y,S_z)$ and $\vec{B}_1(t)$ is the magnetic microwave control field. The environment Hamiltonian $H_E$ consists of the Zeeman and spin-spin interactions for bath spins in the vicinity of the central spin. The system-environment Hamiltonian $H_{SE}$ includes the interactions between the central spin and the bath spins. The bath consists of \car nuclear spins, and unlike electron spins. When the NV is the central spin, the unlike electron spins are the P1 spins. When the P1 is the central spin the unlike electron spins are P1 spins of spectrally-separated $^{14}$N hyperfine manifolds. Since the experiments on the P1 centers in Sample B are performed on the $m_I=0$ hyperfine manifold, the unlike electrons that constitute the bath are those in the $m_I=\pm1$ states.

For a well-quantized central spin (along z), the system-environment Hamiltonian is
\begin{equation}
    H_{SE} = S_z \left[\sum_i\left( A_\perp^iI_x^i + A_{\parallel}^iI_z^i \right) + \sum_{i} J_\parallel^i S_z^i\right].
    \label{Eqn:SysEnvHam}
\end{equation}
The first summation in the brackets is over the nuclear spins, where $A_\perp^i$ and $A_\parallel^i$ are the transverse and longitudinal hyperfine coupling components between the central spin and the $i^\textrm{th}$ \car nuclear spin (with spin operators $I_x^i$ and $I_z^i$). The second summation is over the unlike electron spins in the central spin's environment. $J_\parallel^i$ is the coupling strength between the central spin and the $S_z^i$ component for the $i^\textrm{th}$ unlike electron spin. The transverse components of the electron-electron coupling are non-secular and thus omitted.

In the interaction frame defined by the system and environment Hamiltonians ($H_S + H_E$), the system-environment Hamiltonian $H_{SE}$ becomes time-dependent. 
The system dynamics are approximately generated by an effective central spin Hamiltonian  \cite{yang_quantum_2016}:
\begin{equation}
    H^{(E)}_{SE}(t) = -\gamma_e b_z(t) S_z \ ,
    \label{Eqn:SemiClassNoiseHam}
\end{equation}
where $b_z(t)$ is the $z$-component of the time-varying magnetic field noise at the site of the central spin.   It is the sum of the magnetic fields produced by the fluctuating environment spins and contains periodic components at the \car Larmor frequency. 

For the denser P1 centers in Sample A, the system Hamiltonian in Equation~\ref{eqn:TotHam} also includes dipolar couplings between P1 centers within the same $m_I = 0$ manifold.  As a result the effective ``noise" seen by a single P1 center will no longer look like an effective classical field as described by Equation~\ref{Eqn:SemiClassNoiseHam}. Thus, while we can still perform the analysis along the lines described below, the results need to be interpreted more carefully.

\subsection{The CPMG experiment and noise spectroscopy}

The application of a $\pi$-pulse flips the sign of $H^{(E)}_{SE}$  of Equation \ref{Eqn:SemiClassNoiseHam} in a toggling frame representation. For DD sequences, which consist of a train of $\pi$ pulses, the repeated flipping of the sign can be expressed by a square-wave frame modulation function $f_z(t,\tau)$ \cite{viola_dynamical_1999, cywinski_how_2008, biercuk_dynamical_2011, alvarez_measuring_2011}.  The effective toggling frame Hamiltonian for the dephasing noise acting on the central spin under application of DD is expressed as
\begin{equation}
    \tilde{H}_{SE}(t,\tau) = -\gamma_e f_z(t,\tau) b_z(t) S_z.
\end{equation}

In the CPMG sequence, a series of $\pi_y$ pulses is applied such that the $N^\textrm{th}$ $\pi_y$ pulse occurs at time $t_N = \tau(N-1/2)$ and the $N^\textrm{th}$ echo is detected at time $t = N \tau$.
The amplitude of the $N^\textrm{th}$ echo is given by the expectation value for $S_y$:
\begin{equation}
   \langle S_y  (t,\tau)\rangle = -\frac{\epsilon}{2}\cos(\phi(t,\tau))
\end{equation}
where $\phi(t,\tau)S_z = \int_0^tdt'\tilde{H}_{SE}(t',\tau)$.   The coherence $L(t,\tau)$ is given by the normalized ensemble average of $S_y$:
\begin{equation}
    L(t,\tau) = \frac{\langle S_y (t,\tau) \rangle_\textrm{ens}}{\langle S_y (0,\tau)\rangle_\textrm{ens}} 
    = \langle \cos(\phi (t,\tau)) \rangle_\textrm{ens} 
\end{equation}

For a disordered system composed of many weak couplings to the central spin, it is useful to make the approximation that $b_z(t)$ is drawn from a stationary Gaussian random distribution with zero mean. Then $\phi(t,\tau)$ for the ensemble is a Gaussian random variable \cite{yang_quantum_2016}. 
In this case, the coherence can be shown to be $L(t,\tau) = e^{-\langle \phi(t,\tau)^2 \rangle/2}$, where $\langle \phi(t,\tau)^2 \rangle$ is the variance of the ensemble phase distribution. We set $\langle\phi(t,\tau)^2\rangle/2 = \chi(t,\tau)$ such that the coherence can simply be expressed as $L(t,\tau) = e^{-\chi(t,\tau)}$.

The quantity $\chi$ can be expressed in terms of a dimensionless filter function $F(\omega,t,\tau)$ and the power spectral density $S(\omega)$ of the magnetic noise $b_z(t)$.
\begin{equation}
    \chi(t, \tau)
    = \frac{t^2}{2} \int_{-\infty}^\infty d\omega  \, S(\omega)F(\omega,t,\tau).  
\label{Eqn:ChiOverlapContinuous}
\end{equation}
The filter function $F(\omega,t,\tau)$ of the DD sequence is
\begin{equation}
    F(\omega,t,\tau) = \frac{1}{t^2}\left| \frac{1}{\sqrt{2\pi}} \int_0^{t} f_z(t',\tau) e^{-i\omega t'}  dt' \right|^2,
    \label{Eqn:FilterFromModFunc}
\end{equation}
and the power spectral density $S(\omega)$ is the Fourier transform of $g(t') = \gamma_e^2 \langle b_z(\tilde{t})  b_z(\tilde{t}+t') \rangle$, the autocorrelation function of $b_z(t)$
\begin{equation}
    S(\omega) = \frac{1}{\sqrt{2\pi}}\int_{-\infty}^{\infty} g(t') e^{-i\omega t'} \, dt'.
\end{equation}

The challenge of noise spectroscopy is to calculate $S(\omega)$, given the measured $\chi$ values and the form of the filter $F$ that describes the control sequence. 
In the large $N$ limit of instantaneous $\pi$ pulses, the filter function can be approximated as the sum of $\delta$-functions
\begin{equation}
   F(\omega,t, \tau) \approx \frac{4}{t \tau^2} \sum_{m = \pm \textrm{odd}}^\infty \frac{\delta(\omega - m \pi/\tau)}{\omega^2} \label{Eqn:UsefulFDelta}.
\end{equation}
which results in a discrete sum expression for $\chi$:
\begin{equation}
   \chi(t, \tau) = \frac{4 t}{\pi^2} \sum_{m = 1,3,5,...}^{\infty} \frac{S(\omega_m)}{m^2},
   \label{Eqn:ChiDiscrete}
\end{equation}
where $\omega_m=m \pi/\tau$.
If $S(\omega_1) \gg S(\omega_m)$ for all $m>1$, it is appropriate to approximate $F$ as a single peak at the fundamental ($m=1$) frequency $\omega = \pi/\tau$ and ignore the contributions of higher harmonics, 
\begin{equation}
    S(\omega) = S(\omega_1) = \frac{\pi^2 \ \chi(t, \tau)}{4 t}.
    \label{Eqn:SFundamental}
\end{equation}

For a smooth spectrum that monotonically decays with $\omega$, a sequence with a shorter $\tau$ value decouples the system from a greater range of low-frequency noise and therefore extends coherence more than a sequence with a longer $\tau$ value does. However, if a peak in the filter function $F$ overlaps with a sharp peak in the power spectrum $S$, this will cause a sharp dip in the coherence. Equivalently, if $\tau=mT/2 \, \, (m=1,3,5...)$, where $T$ is the oscillatory period of a prominent noise source, such as the \car Larmor precession, then the coherence decays much more quickly than it otherwise would.  This sensitivity is the key to AC magnetometry.

\begin{figure*}[ht]
\centering
\includegraphics[width=0.9\textwidth]{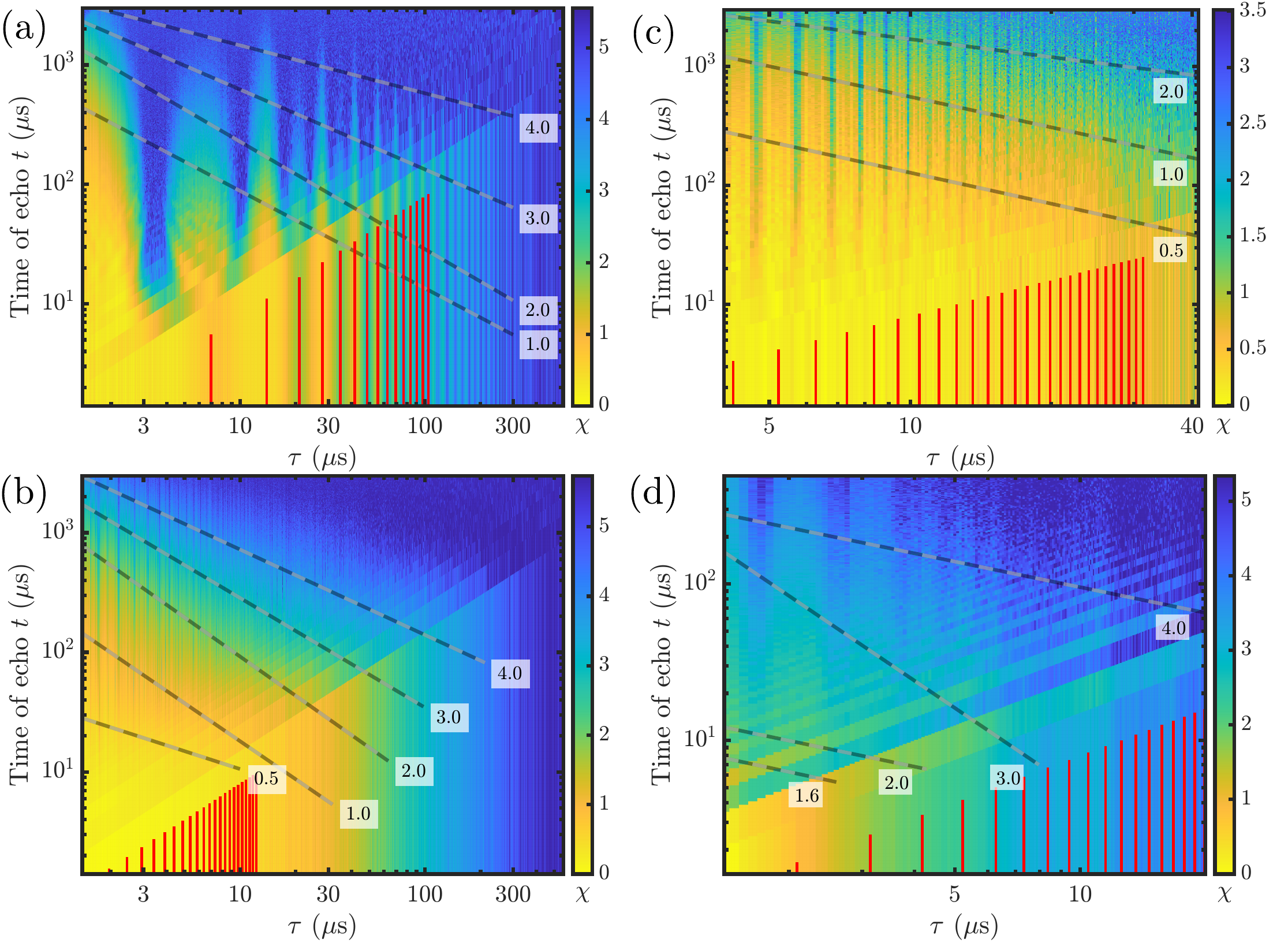}
\caption{\label{fig:2} 
Image plots of $\chi$ data from CPMG experiments. (a) and (b) show data obtained with Sample A on NV centers at 13 mT and 190 mT respectively. (c) and (d) show data obtained with P1 centers at 89 mT in Samples B and A respectively. $\chi=0$ indicates the maximum echo amplitude for the dataset. These data are $T_1$-corrected as described by Equation \ref{Eqn:ChiT1Cor}. The broad character of the power spectrum is captured by the transparent black and white dashed lines, which are power law fits to contours of constant $\chi$ given by Equation \ref{Eqn:tdIfPowerLaw}. The corresponding $\chi$ values are labeled at the lower right ends of the lines. The red vertical lines at the bottoms of the plots indicate $\tau$ values where coherence revivals are expected: $\tau= m \pi/\omega_\textrm{13C}$, where $m$ is even and $\omega_\textrm{13C}$ is the \car Larmor frequency. For the 13 mT in (a), the fits are applied only to the local $t$ maxima of the $\chi$ contours, which occur at these revival $\tau$ values.
}
\end{figure*}

\section{Results and Discussion}
\label{Sec:Results}
\subsection{Overview of data}
\label{Sec:OverviewData}

The observed coherence decay is a product of the decay due to spin bath interactions and, to a lesser extent, $T_1$ relaxation, $L_{\textrm{obs}}(t) = \exp[-(t/(2T_1) - \chi(t)]$. 
In the absence of a spin bath, we expect that the $T_2$ relaxation would be dominated by spin-lattice interactions and $T_2 = 2T_1$. 
Accordingly, to correct for the effects of $T_1$ relaxation, such that $\chi$ only reflects decoherence due to the bath spins, we calculate $\chi$ as
\begin{equation}
\label{Eqn:ChiT1Cor}
    \chi(t) = -\log(L_{\textrm{obs}}(t)) - \frac{t}{2T_1}.
\end{equation}
The measured $T_1$ times are on the order of milliseconds and have little effect on the NV $\chi$ calculation, for which the coherence times are on the order of 100 $\mu$s. The $T_1$ correction is slightly more significant for the P1 center data in Sample B, which achieves coherence times around $T_1/2$ in the short $\tau$ limit.  Details of $T_1$ and $T_2$ experiments and the resulting values are provided in Supplementary Information S1.

Figure \ref{fig:2} shows image plots of $\chi$ data with the $T_1$-correction of Equation \ref{Eqn:ChiT1Cor} for the NV center experiments at 13 mT and 190 mT ((a) and (b)), and P1 center experiments at 89 mT on the CVD and HPHT diamond samples ((c) and (d)). The vertical axes of the image plots indicate the time at which an echo occurs. Therefore, a single-shot experiment with fixed $\tau$ is represented by a vertical line ascending from bottom to top of the image plot. As the echo amplitudes decay, $L$ decreases from 1 to the noise floor, as limited by the signal-to-noise of the experiment, and $\chi$ increases from 0 to some $\chi_\textrm{max}$, which is about 5.5 for the Sample A NV datasets, 5.2 for the Sample A P1 dataset, and about 3.5 for the Sample B P1 dataset. In the lower right areas of the Figure~\ref{fig:2} image plots, one can see sections of data that are uniform in the vertical direction, and the boundaries of the sections are slanted from lower left to upper right. This is because each section corresponds to a particular echo number $N$. The lower rightmost section appears almost as a triangle, which is all $N=1$ (Hahn echo), with the $N=2$ section directly above it, and so on. The image plot is constructed by defining the $\tau$ and $t$ axes first and then populating the 2D array with $\chi$ values from the data set that most nearly match for each point $(\tau,t)$. 

The $\chi$ data exhibit a combination of a broad background and sharp modulations due to electron spin echo envelope modulations (ESEEM) arising from \car hyperfine interactions. Accordingly, our analysis involves characterization of the background spectrum $S_B$ and the sharp peak $S_P$ due to the \car transverse hyperfine interaction. We model the total spectrum as the sum of these contributions:
\begin{equation}
    S(\omega) = S_B(\omega) + S_P(\omega).
    \label{Eqn:SIsSbgSsp}
\end{equation}

The notable difference between the two NV center datasets, Figures~\ref{fig:2} (a) and (b), is that the sharp ESEEM modulations dominate the 13 mT data (a). These modulations are also apparent in the 190 mT data (b) at lower $\tau$ values, but have approximately 15 times shorter periodicity in $\tau$ at the higher field.  The ESEEM modulations prevent a clear measurement of $S_B$. However, this also means that at 13 mT, $S_P$ can be measured directly using Equation \ref{Eqn:SFundamental}. To obtain the background at 13 mT, we analyze the coherence at revival $\tau$ values ($\tau = m\pi/\omega_\textrm{13C}$, where $m=2,4,6...$), which amounts to the standard method of fitting a decay envelope to ESEEM peak revivals \cite{ryan_robust_2010, romach_measuring_2019, bauch_decoherence_2020}.  The decay of the 13 mT CPMG data at revival $\tau$ values (indicated by the vertical red lines) is similar to the decays at 190 mT.   At 190 mT, the \car modulations are more subtle, allowing a clearer characterization of $S_B$, which is then used in a background subtraction step to characterize $S_P$ with higher harmonics of the CPMG filter function.

Comparing the two P1 center datasets, the decay is notably faster for the high [N] sample, Fig. \ref{fig:2}(d), than the low [N] sample, Fig. \ref{fig:2}(c).  The P1 datasets are displayed with different axes and the ESEEM occurs with approximately the same periodicity in both despite a difference in appearance due to different $\tau$ and $t$ ranges interrogated. The high [N] dataset of Fig.~\ref{fig:2}(d) exhibits a prominent rapid decay with $\chi$ going from 0 to $\sim 2$ at early detection times, $t \lesssim 10 \ \mu$s, followed by a much more slowly decaying tail ($2 \lesssim \chi \lesssim 4$) where ESEEM is more visible.  The smaller \car modulations allow a clearer characterization of $S_B$.

Much of our analysis focuses on the structure of fixed $\chi$ contours throughout the data sets.
A commonly used approach, as in References \cite{yuge_measurement_2011, alvarez_measuring_2011, hernandez-gomez_noise_2018}, is to focus on the contour $\chi=1$, which is encountered when $t = T_2$, resulting in Equation \ref{Eqn:SFundamental} taking the form $S(\omega_1) = \pi^2/(4 T_2)$.
We extend the contour tracing approach to the analysis of contours of densely sampled $\chi$ over the available range of values for two reasons. First, we find that this provides a simple way to quantitatively capture the broad character of the coherence data. Second, since larger $\chi$ values generally occur at later detection times, the $\delta$-function approximation of the filter function has greater validity, enabling use of Equations \ref{Eqn:ChiDiscrete} and \ref{Eqn:SFundamental}. This notion is consistent with the observation that longer CPMG trains lead to greater depth enhancement of the ESEEM data \cite{mitrikas_modulation_2015}. Conversely, in the low $N$ limit, the peak in $F$ is wider and is shifted from $\pi/\tau$, as can be seen in Figure \ref{fig:B1ShaCartoon}(d). 

Examples of the $\chi$ contour traces overlaid on a copy of Figure~\ref{fig:2} are shown in Figure S6 of the Supplementary Information. The traces are obtained from the data as follows. For each $\tau$, we convolve the coherence decay $\chi(t)$ with a Gaussian function: $\chi_\textrm{conv}(t) = \int_{-\infty}^{\infty}dt' \ \chi(t) A \exp[-(t - t')^2/(2\sigma^2)]$, with $\sigma=4 \ \mu$s and $A$ being a normalization constant to preserve the magnitude of $\chi$. This smooths out fluctuations for short $\tau$ decays and leaves longer $\tau$ decays unaffected. The $\chi$ contour is defined as the set of points $(t, \tau)$, where $t$ for a given $\tau$ is the time of the first echo for which $\chi_\textrm{conv}$ is greater than the $\chi$ for the contour being traced.

\subsection{Broad spectrum noise characterization}
\label{Sec:DiscussPowLaw}

\begin{figure}[t!]
\centering
\includegraphics[width=0.5\linewidth]{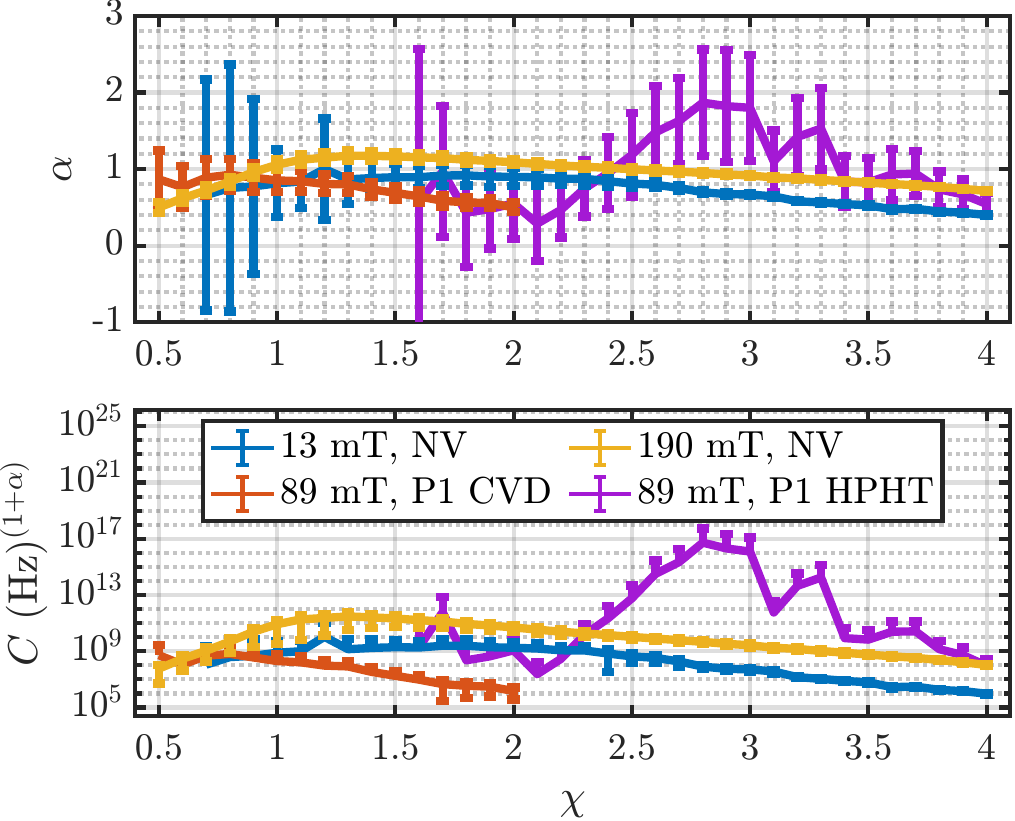}
\caption{\label{fig:3} Parameter values from the $\chi$ contour fits for the power law model of the noise spectrum $S(\omega) = C\omega^{-\alpha}$. The error bars indicate the 95\% confidence interval of the fit. 
See Section \ref{Sec:DiscussPowLaw} for further discussion.
}
\end{figure}

The dominant noise seen by a central NV or P1 center is due to the fluctuations of the surrounding P1 bath.  Thus, the magnetic noise spectrum measured using NS of the NV and P1 centers should reflect similar underlying physics. 
We model the broad spectrum as a power law $S_B(\omega) = C \omega^{-\alpha}$, which enables the inclusion of harmonics that provide a small correction to the calculation of $S_B$ \cite{cywinski_how_2008, alvarez_measuring_2011}. The discrete approximation of Equation \ref{Eqn:ChiDiscrete} becomes
\begin{equation}
    \chi(t, \tau) = C'(\alpha) \, t \, \tau^\alpha,
    \label{Eqn:ChiOfNTauIfPowLaw}
\end{equation}
where
\begin{equation}
    C'(\alpha) = \frac{4 \, \zeta(2+\alpha)}{\pi^{2+\alpha}} \left(1-\frac{1}{2^{2+\alpha}}\right)C,
\end{equation}
with $\zeta$ being the Riemann zeta function. We can find the exponent $\alpha$
by first rearranging Equation~\ref{Eqn:ChiOfNTauIfPowLaw} to the form 
\begin{equation}
    t = \frac{\chi(t, \tau)}{C'(\alpha)\tau^\alpha}.
    \label{Eqn:tdIfPowerLaw}
\end{equation}
If the noise spectrum obeys a power law of the form $S_B(\omega)=C\omega^{-\alpha}$, then, according to Equation \ref{Eqn:tdIfPowerLaw}, a $\chi$ contour should follow a straight line on the $\log(t)$ vs. $\log(\tau)$ axes. The straight black and white dashed lines in Figure \ref{fig:2} are obtained from fitting the contours to Equation \ref{Eqn:tdIfPowerLaw}. The fit parameters $C$ and $\alpha$ from the $\chi$ contours are plotted in Figure \ref{fig:3} and average representative values are provided in Table \ref{Tbl:PowLawParams}. 

\begin{table}[ht]
\caption{\label{Tbl:PowLawParams} Average values of power law ($C\omega^{-\alpha}$) spectrum fit parameters shown in Figure \ref{fig:3}. The row labeled ``$\chi$ range'' indicates the range of representative $\chi$ values used for obtaining these averages.}
\begin{tabular}{|l|l|c|c|c|c|}
\hline
Sample         &  & A      & A      & A      & B      \\ \hline
Species        &  & NV     & NV     & P1     & P1     \\ \hline
$B_0$ (mT)     &  & 190    & 13     & 89     & 89     \\ \hline
$\chi$ range   &  & $0.5-4$  & $1-4$  & $1.6-4$   & $0.5-2$   \\ \hline
$\log_{10}(C)$ &  & 10(1)  & 8(2)   & 11(3)   & 8(1)   \\ \hline
$\alpha$       &  & 0.9(2) & 0.7(2) & 1.0(5) & 0.7(2) \\ \hline
\end{tabular}
\end{table}

While there is some variation in $\alpha$ across the measured $\chi$ values, the data suggests a frequency dependence that scales as $1/\omega^{0.7-1.0}$, as reported in Table \ref{Tbl:PowLawParams}.  
The electronic spin bath surrounding an NV center has been described well by an Ornstein-Uhlenbeck (O-U) process for experiments on single NV spins in samples with low nitrogen concentration \cite{de_lange_universal_2010, bar-gill_suppression_2012, wang_spin_2013, romach_spectroscopy_2015, sung_non-gaussian_2019, sun_self-consistent_2022}. The form of the power spectral density of such a process is a Lorentzian centered at $\omega=0$,
\begin{equation}
    S(\omega) = \frac{2 \Delta^2 \tau_c}{1 + \omega^2 \tau_c^2},
\end{equation}
which has a high-frequency power law tail of the form $1/\omega^2$. 
The origin of the scaling measured in our work being much closer to $1/\omega$ than $1/\omega^2$ is not definitely understood, but we suspect that the high concentration of nitrogen and the reported heterogeneity of the nitrogen concentration throughout the sample play significant roles \cite{bussandri_p1_2023, shimon_large_2022, li_determination_2021, nir-arad_nitrogen_2023}.
One possibility is that the high P1 concentration, which will have much stronger P1-P1 dipolar couplings and exchange couplings \cite{nir-arad_nitrogen_2023}, leads to a fundamentally different type of dynamics and the O-U model no longer applies. Another possibility, motivated by References \cite{dutta_low-frequency_1981, schriefl_decoherence_2006}, is that the high concentration P1 baths still undergo O-U dynamics, but their correlation times $\tau_c$ vary with concentration, giving the appearance of an approximate $1/\omega$ noise in our ensemble measurement. See Supplementary Information S4 for further discussion. The extended CPMG decay tails of the Sample A P1 data, as well as the bi-exponential Hahn Echo decay in Figure S1(e), are highly indicative of the heterogeneity of P1 concentration. Similar heterogeneity effects might explain the roughly $1/\omega$ spectrum measured with the P1 centers in Sample B, however P1 clustering has not been investigated to the same degree in CVD diamonds.

\begin{figure*}[t!]
\includegraphics[width=\textwidth]{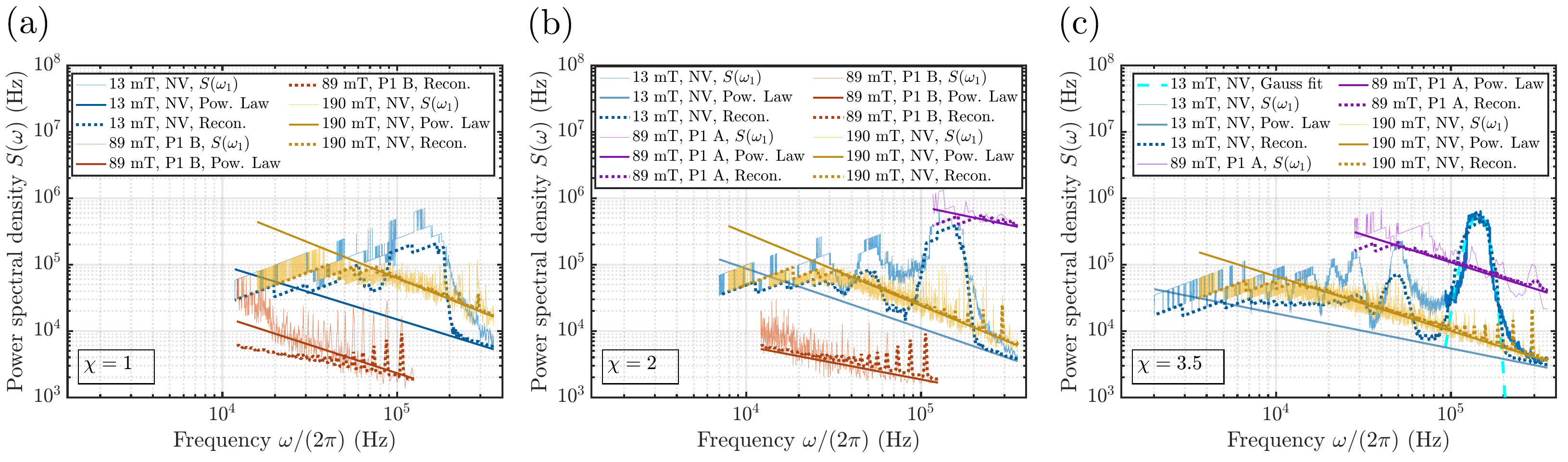}
\caption{\label{fig:4} (a), (b) and (c) all show spectra obtained by using Equation \ref{Eqn:SFundamental} with the $\chi$ contours 1, 2 and 3.5 as indicated. The solid straight lines show the power law fits (``Pow. Law'') obtained using Equation~\ref{Eqn:tdIfPowerLaw}. 
The reconstructed spectra (``Recon.''), which account for the deviation of the $S(\omega_1)$ data from the power law model as $\omega \rightarrow 0$, are discussed in Section \ref{Sec:DiscussPowLaw} and the methods of reconstruction are explained in Supplementary Information S6.A. Due to the contour trace consisting of experimentally realized values of detection times $t$, which are discrete, the low-frequency (large $\tau$) limit of the $S(\omega_1)$ exhibits a jagged appearance, while the high-frequency (short $\tau$) limit appears as more of a continuum. Each of the datasets exhibits subharmonics that arise due to overlap of the CPMG filter harmonics with the \car peak. Only at 13 mT is the \car peak at low enough frequency to be present in the $S(\omega_1)$ spectrum. The cyan dashed line in (c) shows the Gaussian fit to that peak.}
\end{figure*}

Figure~\ref{fig:4} shows the power spectra obtained for $\chi$ = 1, 2 and 3.5 using the power-law model, the approximation in Equation \ref{Eqn:SFundamental}, and a reconstruction technique that accounts for the finite spectrum as $\omega \rightarrow 0$.
We can reliably reconstruct $S(\omega_1)$  from the power law model and a numerically computed CPMG filter function. In this case the power law model is indistinguishable from an alternative spectrum model (such as a stretched Lorentzian) that has a finite value of $S$ at $\omega \rightarrow 0$. The details of the reconstruction are given in Supplementary Information S6.A, and the zero-frequency limit of the power spectra are reported in Supplementary Information S7.

Since the power law fits for the 13 mT dataset were applied to the local maxima of $t$ at revival $\tau$ values, the power laws that appear in the frequency domain follow the minima between the subharmonics of the 13 mT $S(\omega_1)$ spectrum.   The power laws are seen in Figures \ref{fig:4}(a), (b) and (c) to match $S(\omega_1)$ in the mid- to high-frequency region. At lower frequencies, particularly for the 190 mT dataset, the power law is significantly higher than the $S(\omega_1)$ data. 

In Figure \ref{fig:4}, the overall lower level of the spectrum for the P1 centers in Sample B is primarily due to that sample having a much lower concentration of nitrogen \cite{wyk_dependences_1997,bauch_decoherence_2020}. Comparing the 13 mT and 190 mT NV data, we see that the 13 mT spectrum dips lower than the 190 mT spectrum, indicating elevated low-frequency noise at higher field. The cause of this may be increased axial alignment of the P1 bath spins at the higher field. The \car may also play a significant role in this, since at high field, the Zeeman interaction dominates the \car Hamiltonian, but at low field the NV hyperfine interaction is comparable to or dominates the \car Hamiltonian \cite{reinhard_tuning_2012}.

The P1 centers in Sample A are generally in too high of concentration, with strong dipolar couplings, for their dynamics to accurately be captured by the central spin model. Despite this, we apply the analysis to extract an effective power spectrum for the sake of comparison to the other species in this work. The rapid early decay of the Sample A P1 CPMG experiments translates to a higher level of magnetic noise power in the frequency range of dephasing noise as seen in Figure \ref{fig:4}(b) with the $\chi=2$ contour. For $\chi<1.6$ there were not enough data points to obtain a meaningful contour in the Sample A P1 data, and so it is omitted from Figure \ref{fig:4}(a). In the range of $1.6 \lesssim \chi \lesssim 2$, there are very few data points comprising the contour, which leads to very large error bars on the power law fit parameters in Figure \ref{fig:3}. The larger $\chi$ contours, which are traced through the tail of the decays, exhibit power spectra that are more similar to the NV power spectra. This suggests that there is a significant sub-population of P1 centers in Sample A that are more sparsely distributed.

We have also investigated experimentally and numerically how finite pulses and over/under rotation errors affect the power law characterization. These investigations are presented in Supplementary Information S6. We find that the finite pulses in our experiments do not significantly affect the results of the power law characterization. This is because the finite pulses most significantly affect the higher harmonics of the filter function, and in the context of a power law spectrum, these harmonics contribute only a very small amount to the $\chi$ integral in Equation \ref{Eqn:ChiOverlapContinuous}.

\subsection{Characterizing the \car~Larmor frequency}
\label{Sec:Char13C}
The transverse component of the \car hyperfine interaction contributes a peak $S_P$ to the power spectrum at the \car Larmor precession frequency. We can use Equation \ref{Eqn:SFundamental} to observe this peak at 13 mT. However, at 89 mT and 190 mT, the \car Larmor precession period is significantly shorter than the shortest $\tau$ we can use in our CPMG experiments, and we only see the periodic modulation of the coherence curves with increasing $\tau$ as higher harmonics of the filter function overlap with the \car peak. We have developed an analysis procedure that utilizes all applicable harmonics of the filter function to estimate the properties of $S_P$ at high frequencies. The basic approach is outlined below and details are provided in Appendix \ref{Apdx:ScannedHarmonic}.
Note that  $S_P$ could also be generalized to a sum of sharp peaks $\sum_i S_P^i$ for contexts wherein the goal is to obtain the spectrum of a more complex system.

\begin{figure*}[t!]
\includegraphics[width=\linewidth]{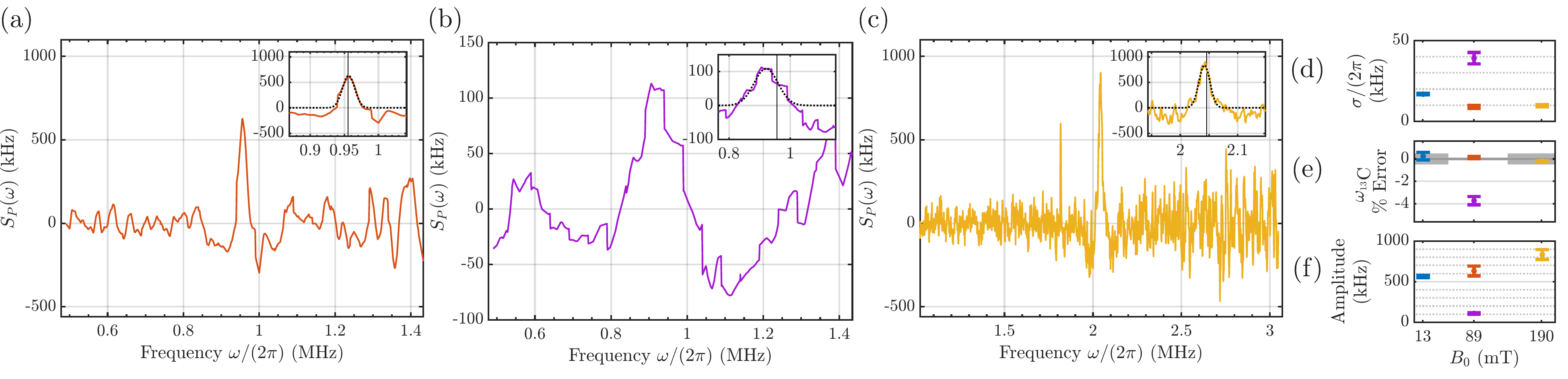}
\caption{\label{fig:5} High-frequency spectra for (a) P1 center in Sample B at 89 mT, (b) P1 center in Sample A at 89 mT, and (c) NV center in Sample A at 190 mT. These spectra are obtained via the harmonic analysis method presented in Appendix \ref{Apdx:ScannedHarmonic}. The insets show detail of the \car peak fit to a Gaussian curve (dotted lines) and the expected \car Larmor frequency (vertical black lines). (d), (e), and (f) show the fit parameters from applying Gaussian fits ($A\exp[-(\omega-\omega_\textrm{13C})^2]/(2\sigma^2)]$) to the \car peak, where $\omega_\textrm{13C}$ is the predicted frequency of the \car Larmor frequency based on experimental conditions. (e) shows the deviation of fit frequency from the predicted frequency, with the gray bands indicating the uncertainty of the prediction. The fits for the 89 mT and 190 mT data are those shown in the insets of (a), (b) and (c). For the 13 mT data, the fit was applied to the $S(\omega_1)$ $\chi=3.5$ spectrum indicated by the cyan dashed line in Figure \ref{fig:4}(c).
}
\end{figure*}

Inserting the composite spectrum expression of Equation \ref{Eqn:SIsSbgSsp} into the discrete $\chi$ calculation of Equation~\ref{Eqn:ChiDiscrete}, we obtain
\begin{equation}
    \chi(t, \tau) =  
    \frac{4t}{\pi^2} \sum_{m=1,3,5...}^\infty \frac{S_B(m\pi/\tau) + S_P(m\pi/\tau)}{m^2}.
\end{equation}
For a sharp peak, we can assume that $S_P(\omega)=0$ for all $m$ except near the sharp feature which is peaked at a frequency $\omega_P$. In the vicinity of the peak we define the scanning frequency $\omega_s$ and scanning harmonic $m_s$ with $\omega_s \equiv m_s \pi/\tau$. Then the summation of $S_P$ collapses to only the $m=m_s$ term:
\begin{equation}
\label{Eqn:ChiAssumeSpSep}
    \chi(t, \tau) = \frac{4t}{\pi^2} 
    \left[ \left(\sum_{m=1,3,5...}^\infty \frac{S_B(m\pi/\tau)}{m^2}\right) 
    + \frac{S_P(\omega_s)}{m_s^2} \right].
\end{equation}
We solve for $S_P$ and express the result as the difference between the measured $\chi$ and the background component $\chi_B$, which was obtained by fitting Equation \ref{Eqn:ChiOfNTauIfPowLaw} as described in Section \ref{Sec:DiscussPowLaw}, resulting in
\begin{equation}
    S_P(\omega_s)
    = \frac{\pi^2 m_s^2}{4 t} 
    \left[\chi(t, \tau) - \chi_B(t,\tau)\right].
    \label{Eqn:SpWithChiDiff}
\end{equation}
Figure \ref{fig:B_ChiSub190Ex} of the Appendix depicts an example of this calculation. When we perform this calculation in our analysis, we include an additional term in the pre-factor to account for finite pulse widths as explained in Supplementary Information S6. For a given combination of $\tau$ and $t$, we first compute $\chi_B$ by determining which power law model (corresponding to the different $\chi$ contour fits) most nearly approaches the given $\tau$ and $t$ point. This is the same as determining by an image plot in Figure \ref{fig:2}, which black and white dashed line most nearly crosses a particular ($t, \tau$) point. That power law model is then used to compute the expected value of $\chi_B$ using Equation \ref{Eqn:ChiOfNTauIfPowLaw}.

A spectrum $S_P$ is obtained by combining the results of evaluating Equation \ref{Eqn:SpWithChiDiff} for all applicable combinations of $\tau$ and $t$ in the dataset. The procedure for this calculation is provided in Appendix \ref{Apdx:ScannedHarmonic}. 
The resulting spectra for the 89 mT and 190 mT data are shown in Figure \ref{fig:5}(a), (b) and (c). We note that while negative values are not permitted in a power spectrum, they appear here because obtaining the spectra $S_P$ via Equation \ref{Eqn:SpWithChiDiff} is effectively a form of baseline subtraction.

Figures \ref{fig:5}(d), (e) and (f) show the fit parameters obtained from fitting a Gaussian to the reconstructed \car spectral line. The error bars indicate the 95\% confidence intervals. For the 13 mT experiment, we use the $\chi=3.5$ contour to obtain the fit. The data to which the fit is applied is the portion of $S(\omega_1)$ between 90 kHz and 200 kHz (indicated by the thicker blue line), and the fit curve is the cyan dashed line in Figure~\ref{fig:4}(c). For the 89 mT and 190 mT analyses, the fit is shown in the insets of Figures \ref{fig:5}(a), (b) and (c). 

The Sample A P1 center spectrum of the \car stands out as having approximately 5x lower amplitude and 5x greater width than NV and Sample B P1 spectra, which are in fairly close agreement and agree well with similar measurements in the literature \cite{hernandez-gomez_noise_2018,  romach_measuring_2019}. Considering that both samples have natural isotopic abundance of \car and that there should be a generally identical distribution of magnetic environments for P1s and NVs in Sample A, the discrepancy indicates that the strong dipolar coupling of P1 centers in Sample A distorts the intrinsic characteristics of the spectrum. In other words, most of the P1 centers in Sample A have such strong dipolar couplings that the \car peak cannot be measured with conventional CPMG. Still though, the signature of \car is evident in the $\chi$ data for P1 Sample A in Figure \ref{fig:2}(d), particularly at later $t$. This exemplifies that the long-lived coherences that follow the rapid initial decay of many-spin systems \cite{hahn_long-lived_2021} can sense peaks in the power spectrum. Note that the precision of the peak measurement is limited by requiring short $\tau$ values to obtain such coherences.

A detailed discussion of the variations in the measured properties of the \car peak is presented in Supplementary Information S5.

\section{Summary and outlook}
\label{Sec:Summary}

The use of ensemble pEPR NS using stroboscopic, inductively-detected measurements will expand our ability to study the properties of optically-dark spins that frequently form the bath for localized spin qubits.  Here we performed DD NS of both P1 and NV centers in diamond, allowing us to directly contrast the local magnetic environments seen by these electronic spin systems.  In the HPHT samples, the high concentration of P1 centers leads to a breakdown of the central spin model when measuring the P1 spins, though it remains valid when characterizing the dynamics of the NV centers. 

All spectral reconstructions showed two prominent features --  a broad background exhibiting a frequency dependence best approximated as $1/\omega^{0.7-1.0}$ -- and a narrow peak caused by the \car transverse nuclear hyperfine interaction.  For the HPHT sample, we suspect that the deviation of this power-law spectrum from the expected O-U dynamics (i.e., Lorentzian spectrum with $1/\omega^2$ tail) is due to the high concentration and heterogeneity of P1 centers in the sample, which is supported by recent studies on clustering \cite{bussandri_p1_2023, shimon_large_2022, li_determination_2021, nir-arad_nitrogen_2023}. It is unclear if this is also the case in CVD diamond.  The heterogeneity of the HPHT diamond also manifested in the presence of multiple decay timescales of the observed coherence in the P1 experiments. 
Finite pulse effects and flip angle errors do not greatly affect the broad spectrum characterization, but can have non-negligible effects on the characterization of sharp peaks in the power spectrum. Finite pulses make CPMG less sensitive to peaks at higher frequencies while flip angle errors can cause an asymmetric splitting of a spectral peak.

The harmonic analysis allowed us to perform more precise measurements of the \car frequency for P1 centers in CVD diamond at 89 mT and NV centers in HPHT diamond at 190 mT than could be obtained with the NV centers at 13 mT experiment. The P1 centers in the HPHT diamond had too short of coherence times to obtain as precise of a measurement, however this many-spin system exhibited weak coherences that persisted to late detection times where detection of the \car peak in the power spectrum was still achieved.

The methods presented in this work can be applied generally when CPMG is used to characterize noise.  These methods applied to diamond could very directly inform strategies for quantum sensing, e.g., for performing nanoscale NMR \cite{schwartz_blueprint_2019} with electron spins as proxy sensors.  Stroboscopic measurements of ensembles opens the door to adaptive measurement strategies such as actively varying $\tau$ to follow a particular $\chi$ contour, or using closed-loop feedback to perform adaptive quantum sensing.

\section{Acknowledgements}  We thank Lorenza Viola, Bhargava Thyagarajan, Linta Joseph, and James Logan for helpful discussions. We thank Lihuang Zhu for helping design the pEPR spectrometer. We are grateful to Kevin Villegas Rosales and Yoav Romach of Quantum Machines for helping set up and program the OPX as well as helpful discussions. We thank Johan van Tol for help with diamond sample characterization that was performed at the National High Magnetic Field Laboratory in Tallahassee, FL. We thank Gajadhar Joshi and Jonathan Friedman for helpful insight on the loop-gap resonator design. We thank Dwayne Adams and Christopher Grant for help building the EPR probe and resonator. This work was partially supported by funding from the National Science Foundation under grant CHE-2203681 and cooperative agreement OIA-1921199 and by the Gordon and Betty Moore Foundation by grant GBMF12251.  We acknowledge support of a QISE-NET Triplets Award with EQW funded by NSF award DMR-1747426.

\newpage

\appendix

\section{Harmonic Analysis}
\label{Apdx:ScannedHarmonic}

Noise spectroscopy strategies such as approximating the filter function as its fundamental peak (Equation~\ref{Eqn:SFundamental}) to compute $S_P(\omega)$ or using Álvarez and Suter's matrix inversion method \cite{alvarez_measuring_2011} enable one to measure the power spectral density up to the frequency $\omega= \pi/\tau_\textrm{min}$, where $\tau_\textrm{min}$ is the shortest $\tau$ one can implement with their experiment hardware. In applications of ac magnetic field sensing, it may occur that there are peaks in the power spectrum at frequencies a few times greater than $\pi/\tau_\textrm{min}$, whose effects can still be seen in experiments in which $\tau$ is swept with a small step size. It is desirable to obtain a wide spectrum that includes these peaks, so that the peaks can be measured with sufficient contrast to the background. Frequency comb \cite{norris_qubit_2016} and narrowband Slepian modulation techniques \cite{frey_simultaneous_2020} are two possible ways to overcome the $\pi/\tau_\mathrm{min}$ limit. Higher harmonics of CPMG-type filters have also been used to estimate the magnitude of a spectral peak at a particular frequency \cite{hernandez-gomez_noise_2018}. Here we present an analysis method for obtaining a wide spectrum that reveals prominent features at frequencies greater than $\pi/\tau_\textrm{min}$.

\begin{figure}[b]
\includegraphics[width=\linewidth]{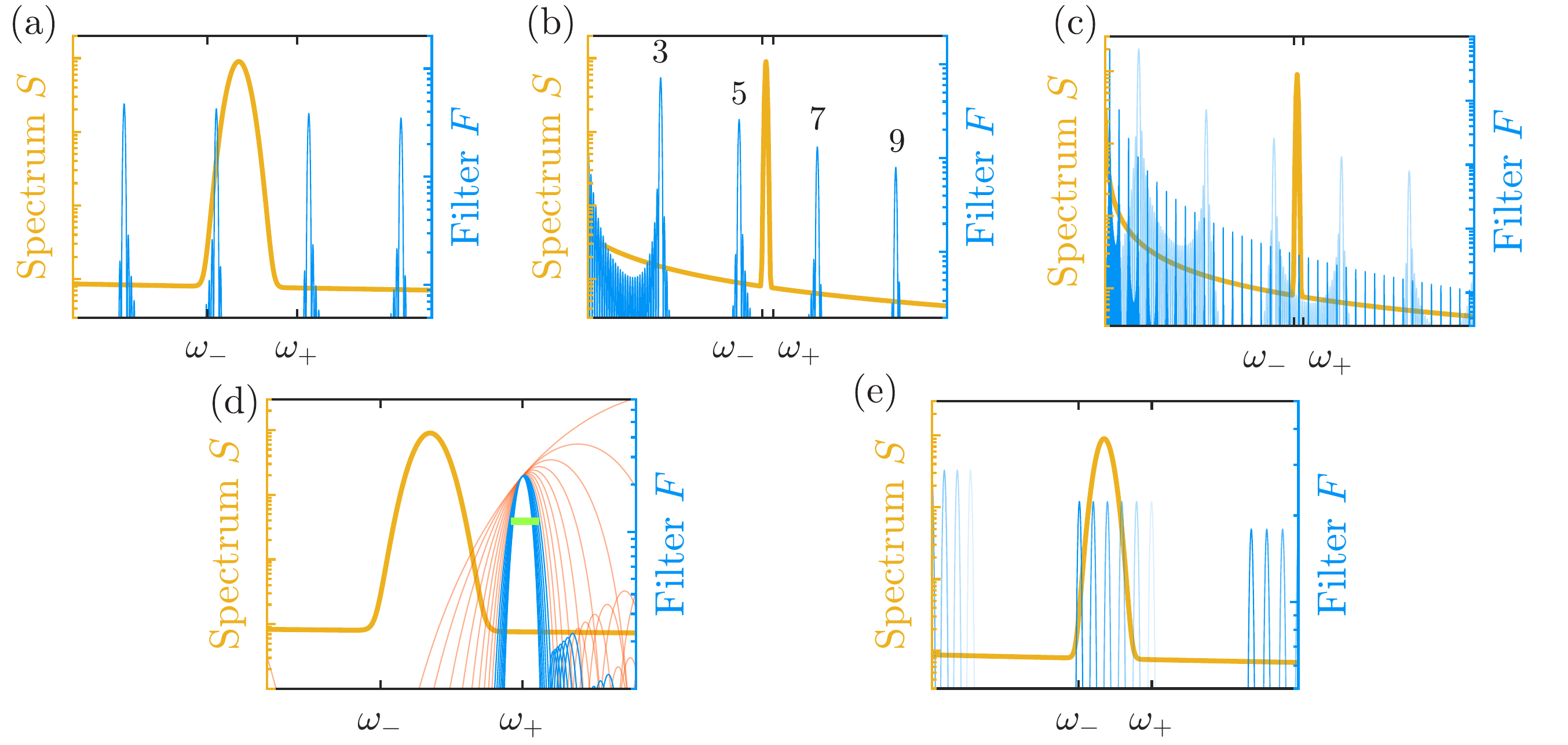}
\caption{\label{fig:B1ShaCartoon} Illustrations of filter considerations in the harmonic analysis. See text for details.}
\end{figure}
 
In essence, the harmonic analysis breaks the range of high frequencies into a series of blocks of width $W$, which are instances of a scanning window. For each instance of the window, Equation \ref{Eqn:SpWithChiDiff} is computed for each valid combination of $\tau$ and echo number $N$ for which a harmonic $m_s$ of the filter function is present within the window. Each calculation gives a point in the spectrum at frequency $\omega = m_s \pi/\tau$. Interpolation and averaging are used to convert the set of calculated spectral points into one continuous spectrum. Results from this method are presented in Figures \ref{fig:5}(a), (b) and (c) and S10(g) and (h).

In our case, the shortest interpulse delay $\tau$ that we can apply is 1.4 $\mu$s, as explained in Supplementary Information S3. The fundamental ($m=1$) frequency of a CPMG filter function is $\omega=\pi/\tau$. So the highest frequency that we can fundamentally measure is $\omega \approx 360 \times 2\pi$ rad kHz, which can be seen as the maximum frequency of the spectra in Figures \ref{fig:4}(a), (b) and (c) and S10(e). In those spectra however, the subharmonics of the \car peak are clearly visible. These subharmonics, which correspond to CPMG sequences with faster decays, occur at frequencies for which a higher harmonic of the filter function overlaps with the \car peak, i.e., when $\omega=\omega_\textrm{13C}/m$, where $\omega_{13\textrm{C}}$ is the \car nuclear Larmor frequency and $m$ is an odd integer.

\begin{figure}[t!]
\centering
\includegraphics[width=0.7\linewidth]{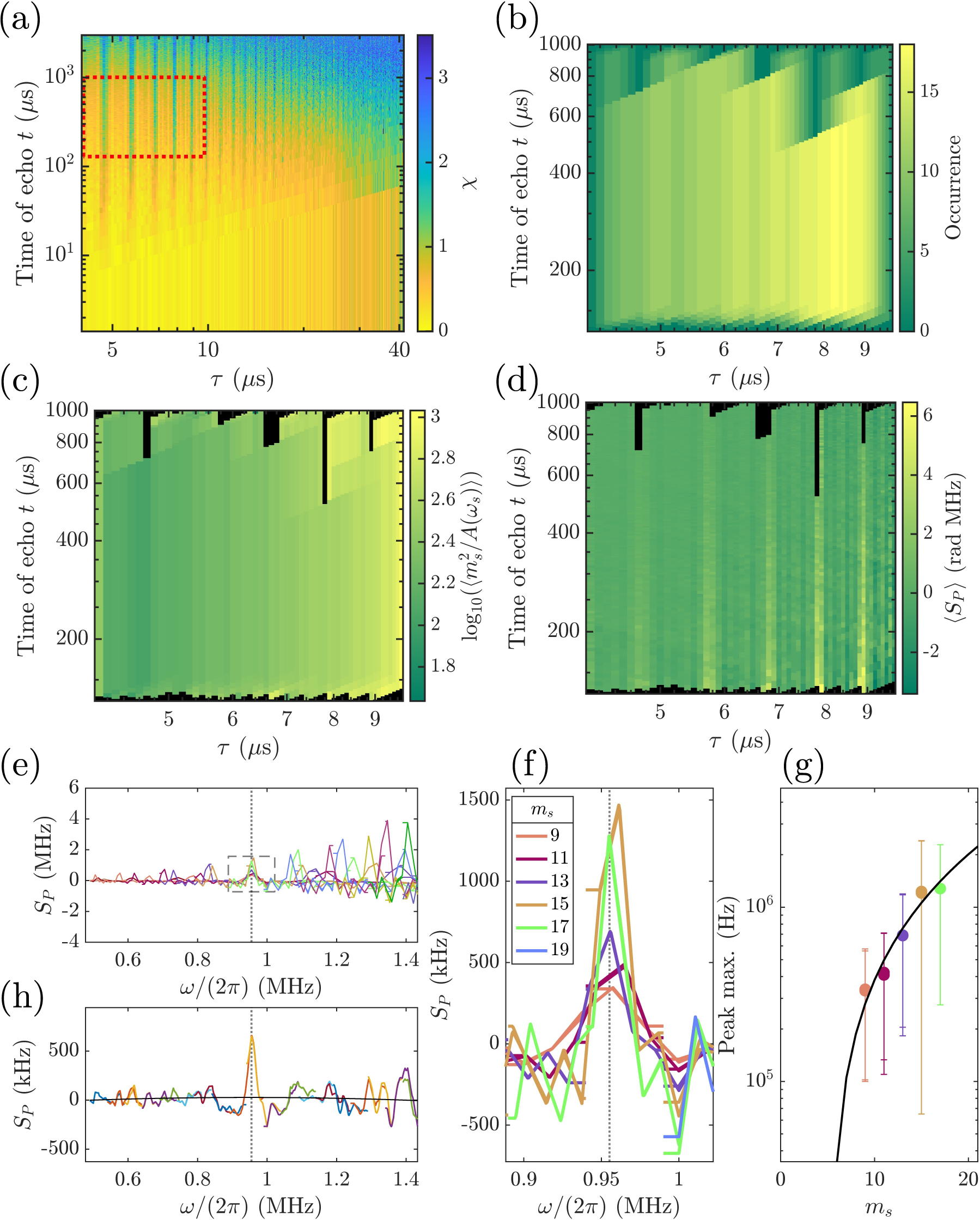}
\caption{\label{fig:B2Process090} Visualizations of parts of the harmonic analysis procedure for the 89 mT P1 Sample B data. (a) Image plot of the coherence data. The red dashed rectangle indicates the ``sensing region'' over which it is valid to use the $\delta$ approximation of the filter function to characterize the \car peak. This is the region that plots (b), (c) and (d) show. (b) Number of times each point of the data is used in the harmonic analysis. (c) Log$_{10}$ of the average harmonic weighting prefactor in the calculation of Equation \ref{Eqn:SpWithChiDiff}. $A(\omega_s)$ is the finite filter correction explained in Supplementary Information S6.A. (d) Average contribution of each data point in the calculation of $S_P$ in the harmonic analysis. (e) Spectral segments described in Steps 5 and 6 of the harmonic analysis. A segment is obtained using one harmonic $m_s$ and averaged over all available $N$. The vertical gray dotted line represents the \car Larmor frequency. (f) Detail of the area outlined by the gray dashed box in (e) with the legend indicating which filter harmonic each segment was obtained with. (g) Maxima of the peaks shown in (f). The black line is a quadratic fit. See Supplementary Information S5 for discussion. (h) Overlapping segments of width $W$ obtained when averaging the contributions from all available harmonics as discussed in Steps 5 and 6 of the harmonic analysis. These are then averaged to produce the spectrum in Figure \ref{fig:5}(a).}
\end{figure}

\begin{figure}[t!]
\centering
\includegraphics[width=0.7\linewidth]{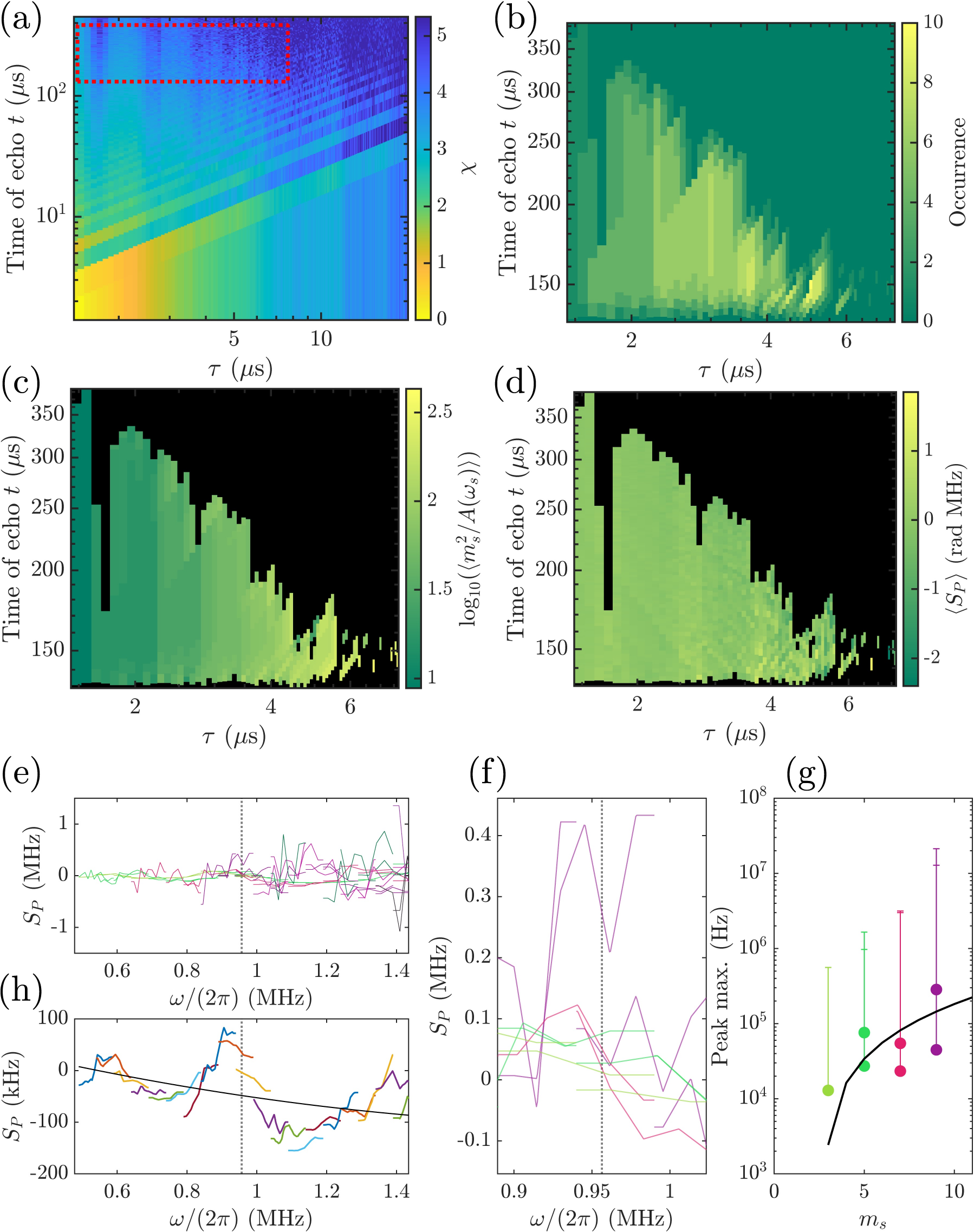}
\caption{\label{fig:B4ProcessP1Hpht} Visualizations of parts of the harmonic analysis procedure for the 89 mT P1 Sample A data. See the caption to Figure \ref{fig:B2Process090} for descriptions. The averaged segments in (h) are averaged to produce the spectrum in Figure \ref{fig:5}(b).}
\end{figure}

\begin{figure}[t!]
\centering
\includegraphics[width=0.7\linewidth]{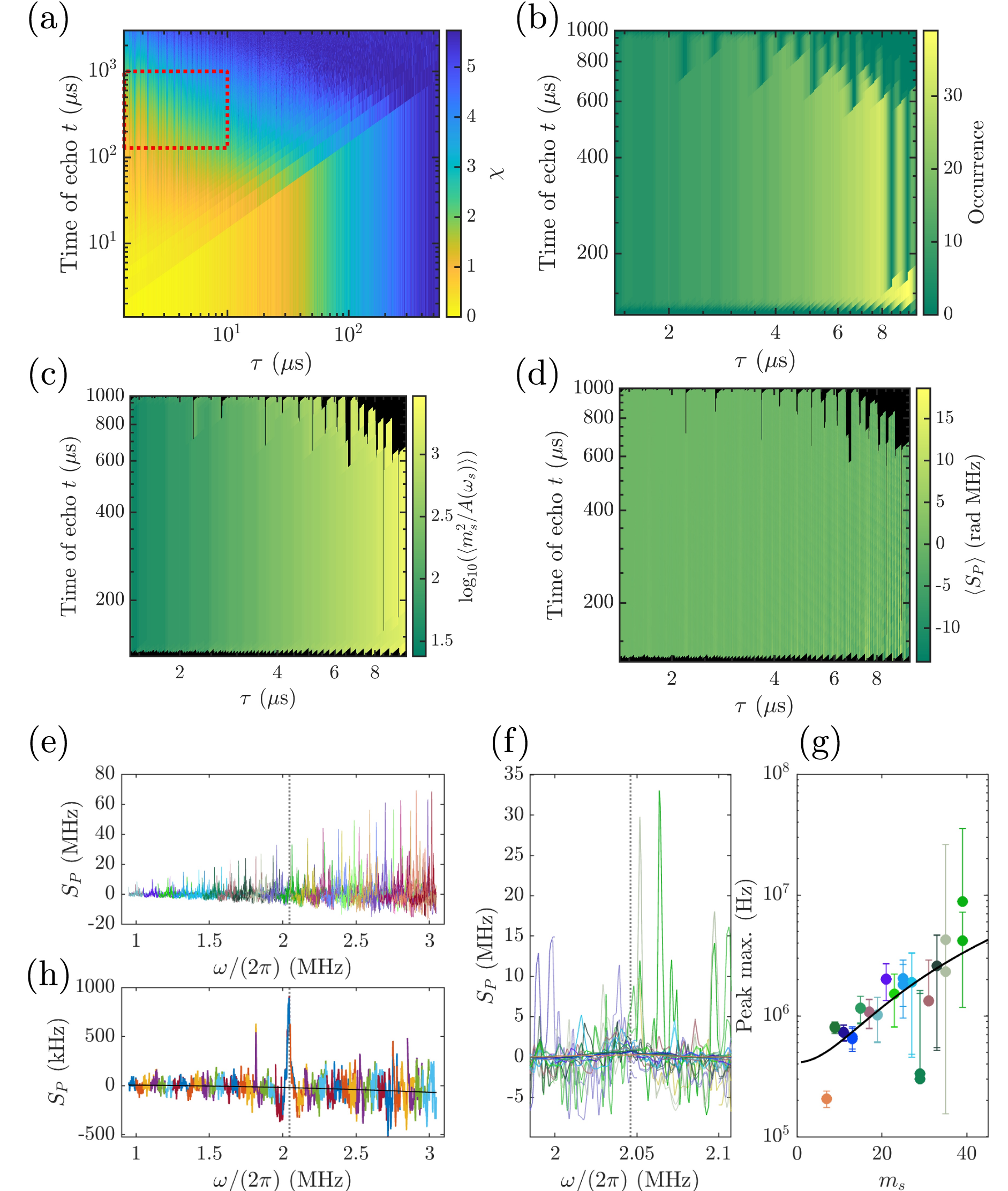}
\caption{\label{fig:B3Process190} Visualizations of parts of the harmonic analysis procedure for the 190 mT NV data. See the caption to Figure \ref{fig:B2Process090} for descriptions. (f) does not include a legend as in Figure \ref{fig:B2Process090} because the spectral segments are too dense to separately distinguish. The averaged segments in (h) are averaged to produce the spectrum in Figure \ref{fig:5}(c).}
\end{figure}

The \textit{valid} subharmonics are ones which meet certain criteria for avoiding sensitivity limitations, avoiding distortions caused by finite widths of the peaks in the filter function, and avoiding distortions caused by multiple filter peaks overlapping with a single spectral line. These three criteria define a maximum and minimum detection time $t$ and a maximum $\tau$ respectively. These limits can be thought of as defining a sensing region. For the 89 mT and 190 mT data analyzed in this work, the sensing regions are those indicated by the red dashed rectangles in Figures \ref{fig:B2Process090}(a), \ref{fig:B4ProcessP1Hpht}(a) and \ref{fig:B3Process190}(a). The criteria for avoiding sensitivity and distortion limitations inform the following 6-step analysis protocol. Steps 1 through 5 explain the frequency scan procedure over one instance of the window, focusing on an example where the window contains a prominent peak, as depicted by the cartoon examples in Figure \ref{fig:B1ShaCartoon}. Step 6 explains how to iterate this procedure to obtain a wide spectrum.

\begin{figure}[ht]
\centering
\includegraphics[width=0.5\linewidth]{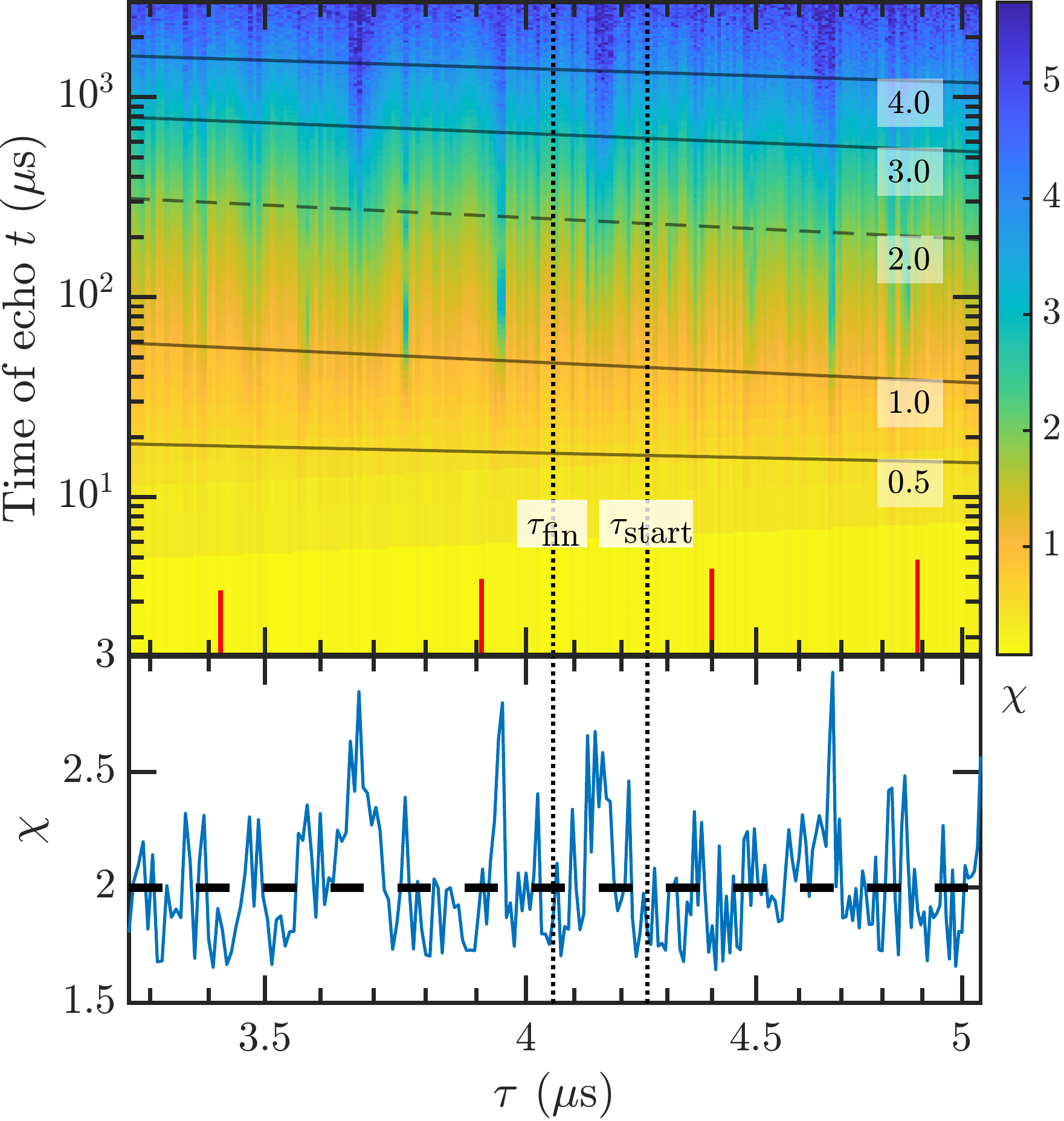}
\caption{\label{fig:B_ChiSub190Ex} Example of the background subtraction (Equation \ref{Eqn:SpWithChiDiff}) in the harmonic analysis. The top plot shows a zoomed in section of the image coherence plot of the 190 mT NV data shown in Figure \ref{fig:2}(b). The bottom plot shows the value of $\chi$ along the $\chi=2$ power law fit line (dashed line). This figure serves as a visual example of the \car peak $S_P$ calculation described in Step 5 of the harmonic analysis.}
\end{figure}

1. Determine the width $W$ of the frequency-scanning window. $W$ is determined by the requirement that only one harmonic of the filter function overlap with the peak. This is the assumption that allows the isolation of $S_P$ from the summation in Equation \ref{Eqn:ChiAssumeSpSep}. $W$ is the maximum peak width that can be resolved by the analysis process. $W$ sets the maximum $\tau$ value that will contribute to the calculation as $\tau_{\textrm{max}} = 2\pi/W$. The center of the scan window is denoted $\omega_c$, and the upper $(+)$ and lower $(-)$ bounds of the window are $\omega_{\pm} = \omega_c \pm W/2$. With $\tau \leq \tau_{\textrm{max}}$, the minimum frequency difference between two consecutive peaks of the filter function is greater than or equal to $W$, as depicted in Figure \ref{fig:B1ShaCartoon}(a). We note that in some systems $\tau_\textrm{max}$ may be limited by a hardware or coherence limit, in which case $W$ should be defined from $\tau_\textrm{max}$.

2. Determine the range of harmonics, i.e., peaks of the filter $F$, that can be used as the scanning harmonic $m_s$ (where $m_s$ can be odd integers in reference to $m$ of Equation \ref{Eqn:ChiDiscrete}). For $m_{\textrm{min}}$ we want the lowest filter peak available. With the shortest $\tau$ that can be applied with the hardware, $m_\textrm{min}$ is the lowest peak that can still reach $\omega_+$. For the example in Figure \ref{fig:B1ShaCartoon}(b), which shows the filter for CPMG with $\tau_\textrm{min}$, we get $m_{\textrm{min}} = 7$ because the $m=5$ peak falls short of $\omega_+$. Formally, $m_{\textrm{min}}$ is the lowest odd integer greater than $\omega_{+}\tau_{\textrm{min}}/\pi$. As $\tau$ increases, the peaks in $F$ contract toward 0. The maximum harmonic $m_{\textrm{max}}$ is then determined by requiring that only one harmonic can be in the range $[\omega_-,\omega_+]$. Specifically, when the $m_\textrm{max}$ peak is at $\omega_-=\pi m_\textrm{max}/\tau_\textrm{max}$, the $m_\textrm{max}+2$ peak should be at a frequency greater than $\omega_+$:
\begin{equation}
    \frac{(m_\textrm{max}+2)\pi}{\tau_\textrm{max}} \geq \omega_{-} + W.
\end{equation}
This inequality is more succinctly expressed in terms of $W$. We say that $m_\textrm{max}$ is the highest odd integer subject to the constraint
\begin{equation}
    m_\textrm{max} \leq \frac{2 \omega_-}{W}.
\end{equation}
Figure \ref{fig:B1ShaCartoon}(c) shows the filters for $\tau_\textrm{min}$ (light blue) and $\tau_\textrm{max}$ (darker blue).

3. For a given scan harmonic $m_s$ determine the starting and finishing $\tau$ values to iterate over. Assuming we are sweeping $m_s$ in the direction of increasing $\omega$ (as inferred by the increasing transparency of the blue peaks in Figure \ref{fig:B1ShaCartoon}(e)), then the $\tau$ values are swept in decreasing order, $\tau_\textrm{start} > \tau_\textrm{fin}$ (see the example in Figure \ref{fig:B_ChiSub190Ex}). The corresponding frequencies $\omega_\textrm{start} = m_s \pi/\tau_\textrm{start}$ and $\omega_\textrm{fin} = m_s \pi/\tau_\textrm{fin}$ will not necessarily be equal to $\omega_-$ and $\omega_+$. Rather, $\omega_\textrm{start}$ and $\omega_\textrm{fin}$ should be within the bounds of $\omega_-$ and $\omega_+$.
\begin{eqnarray}
    \omega_\textrm{start} &\geq \omega_- \\
    \frac{m_s \pi}{\tau_\textrm{start}} &\geq \omega_- \\
    \tau_\textrm{start} &\leq \frac{m_s \pi}{\omega_-}.
\end{eqnarray}
$\tau_\textrm{start}$ is the highest available $\tau$ subject to this constraint. Similarly,
\begin{eqnarray}
    \omega_\textrm{fin} &\leq \omega_+ \\
    \frac{m_s \pi}{\tau_\textrm{fin}} &\leq \omega_+ \\
    \tau_\textrm{fin} &\geq \frac{\pi m_s}{\omega_+}.
\end{eqnarray}
$\tau_\textrm{fin}$ is the lowest available $\tau$ subject to this constraint.

4. For a given $m_s$ and associated range $[\tau_\textrm{fin},\tau_\textrm{start}]$, determine the range of echo numbers $[N_\textrm{min},N_\textrm{max}]$. To determine $N_\textrm{min}$, it must be taken into account that the peaks in the filter function $F$ have finite width, which becomes narrower with increasing $N$. Therefore, $N_\textrm{min}$ should be set by requiring that the harmonic peak is narrower than the sharp peak needing to be resolved. Without knowing the sharp peak's width beforehand, one can instead require that the harmonic peak width be much narrower than the scan frequency range $W$. With $\sigma = \sqrt{2 \pi}/(N \tau)$ being the width of the scan harmonic peak (obtained from the Gaussian approximation of the peaks in the filter $F$ \cite{yang_quantum_2016}), we require $\sigma < \epsilon W$, with $\epsilon \ll 1$. This leads to
\begin{equation}
    N_\textrm{min} > \frac{\sqrt{2\pi}}{\epsilon W \tau_\textrm{fin}}.
\end{equation}
$N_\textrm{min}$ is the lowest $N$ available subject to this constraint. In Figure \ref{fig:B1ShaCartoon}(d), filters are shown for a fixed $\tau$ and $N=1$ through 16. The horizontal green bar has width $2\epsilon W$. The red filters are $1 \leq N \leq 9$ and fail the width criterion. The blue filters are $10\leq N \leq 16$ and satisfy the width criterion.
$N_\textrm{max}$ should be determined by some SNR threshold or $T_1$ limit. We use $N_\textrm{max}\tau_\textrm{start} < 1$ ms, which is sufficiently shorter than the 2 ms $T_1$ of the P1 center. Note, in the definitions of $N_\textrm{min}$ and $N_\textrm{max}$, we have inserted $\tau_\textrm{fin}$ and $\tau_\textrm{start}$ respectively, to place the more restrictive constraints on the $N$ range. This ensures that the $N$ range meets the requirements for the entire $\tau$ range and the given $m_s$. Since the detection time is $t = N \tau,$ these width limits set a maximum and minimum $t$, which are the tops and bottoms of the red dashed rectangles in Figures \ref{fig:B2Process090}(a) and \ref{fig:B3Process190}(a).

5. Compute $S_P(m_s \pi/\tau)$ using Equation \ref{Eqn:SpWithChiDiff}. For each combination of $m_s$ and $N$, a segment of the spectrum $S_P(\omega)$ is obtained by sweeping $\tau$ from $\tau_\textrm{start}$ to $\tau_\textrm{fin}$. These segments are averaged over the available $N$ (seen in (e) and (f) of Figures \ref{fig:B2Process090}, \ref{fig:B4ProcessP1Hpht} and \ref{fig:B3Process190}). Then each of those segments is averaged over the available $m_s$ resulting in a single segment seen in (h) of Figures \ref{fig:B2Process090}, \ref{fig:B4ProcessP1Hpht} and \ref{fig:B3Process190}.  Figure \ref{fig:B_ChiSub190Ex} provides example data from the 190 mT experiment to illustrate how Equation \ref{Eqn:SpWithChiDiff} is calculated. The effective bounds of the scanning window represented in the $\tau$-domain by $\tau_\textrm{start}$ and $\tau_\textrm{fin}$ are indicated by the vertical dotted lines. In the middle of this window, we see more rapid decay due to a filter harmonic overlapping with the \car peak. $\chi(t, \tau)$ is the exact value of $\chi$ at the point in question. To determine $\chi_B$ for a given combination of $\tau$ and $N$, we determine which $\chi$ contour fit (straight black lines) most closely crosses over that point, and use the parameters from that fit to calculate the precise value of $\chi_B$. We solve Equation \ref{Eqn:ChiOfNTauIfPowLaw} for the various $\chi$ contour characterizations, and find which $\chi$ contour fit predicts the closest $N$ to the point in question with the given $\tau$.
For example, the point at $\tau = 4.15 \ \mu$s and $t = N \tau = 230 \ \mu$s has the $\chi=2$ contour fit (dashed line) crossing over it, so $\chi_B = 2$ for that point. The lower plot in Figure \ref{fig:B_ChiSub190Ex} shows the value of the measured $\chi$ along the power law fit of the $\chi=2$ contour. So the subtraction term in brackets of Equation \ref{Eqn:SpWithChiDiff} is $\chi - \chi_B = 2.5 - 2 = 0.5$. In experiments, it may be the case that $\tau$ is sampled linearly with hardware-limited precision. Therefore, the spectrum of $S_P(\omega)$ has $1/\omega$ sampling. The array of $S_P$ values can be resampled, using interpolation, onto a linearly spaced array of $\omega$ values with a small step size to create a segment of the spectrum. The linearly resampled spectrum segments can then be added together for each combination of $m_s$ and $\tau$ for averaging.

6. Perform Steps 1-5 for a wide sweep of the window position $\omega_c$, while maintaining the fixed window width $W$. The spectral segments of adjacent ranges $[\omega_-,\omega_+]$ can then be stitched together in order to generate the spectrum over a wide frequency range. Furthermore, to average out interpolation artifacts that may arise at the edges of the scanning window, one can overlap the ranges such that any range starts at $W/2$ greater than the $\omega_-$ of the previous range as in a bricklaying pattern as in (h) of Figures \ref{fig:B2Process090}, \ref{fig:B4ProcessP1Hpht} and \ref{fig:B3Process190}.

Figures \ref{fig:B2Process090}, \ref{fig:B4ProcessP1Hpht} and \ref{fig:B3Process190} show intermediate steps of the harmonic analysis for the P1 data at 89 mT and NV data 190 mT data. The plots (a) of Figures \ref{fig:B2Process090}, \ref{fig:B4ProcessP1Hpht} and \ref{fig:B3Process190} indicate the valid sensing region with the red dashed rectangles. We set the maximum detection time $t$ to be 1 ms for the P1 Sample A and NV data, and 450 $\mu$s for the P1 Sample B data, to stay well within $T_1$ relaxation limitations. We set the maximum $\tau_\textrm{max} \leq 10 \ \mu$s, which gives $W = 2\pi\times 100$ rad kHz, approximately 5 times wider than the width ($2\sigma$) of the \car peak. We set $W\epsilon = 2\pi \times 3$ kHz, which is sufficiently narrower than the \car peak and corresponds to a minimum detection time of $t \approx 133 \ \mu$s.
Figures (b), (c) and (d) of \ref{fig:B2Process090}, \ref{fig:B4ProcessP1Hpht} and \ref{fig:B3Process190} reveal how each data point in the sensing region is incorporated into the harmonic analysis. The (b) figures simply show how many times each point was used in the calculation of Equation \ref{Eqn:SpWithChiDiff}. In order to avoid the sensitivity limits of decoherence, points were omitted if the $\chi$ value was greater than a noise threshold: 2 for the Sample B data and 4.5 for the Sample A data.

\newpage

\section*{References}
\bibliographystyle{unsrt}
\bibliography{main}

\end{document}


\title[Supplementary information: Characterizing the magnetic noise power spectrum ...]{Supplementary information: Characterizing the magnetic noise power spectrum of dark spins in diamond}

\author{Ethan Q. Williams}
\ead{ethan.q.williams.gr@dartmouth.edu}
\author{Chandrasekhar Ramanathan}
\ead{chandrasekhar.ramanathan@dartmouth.edu}

\vspace{10pt}

\address{Department of Physics and Astronomy, Dartmouth College, Hanover, New Hampshire 03755, USA}


\vspace{10pt}

\begin{indented}
\item[]August 2024
\end{indented}


\section{Diamond samples}
\label{sisec:Samples}

\begin{figure}[ht]
\includegraphics[width=0.75\linewidth]{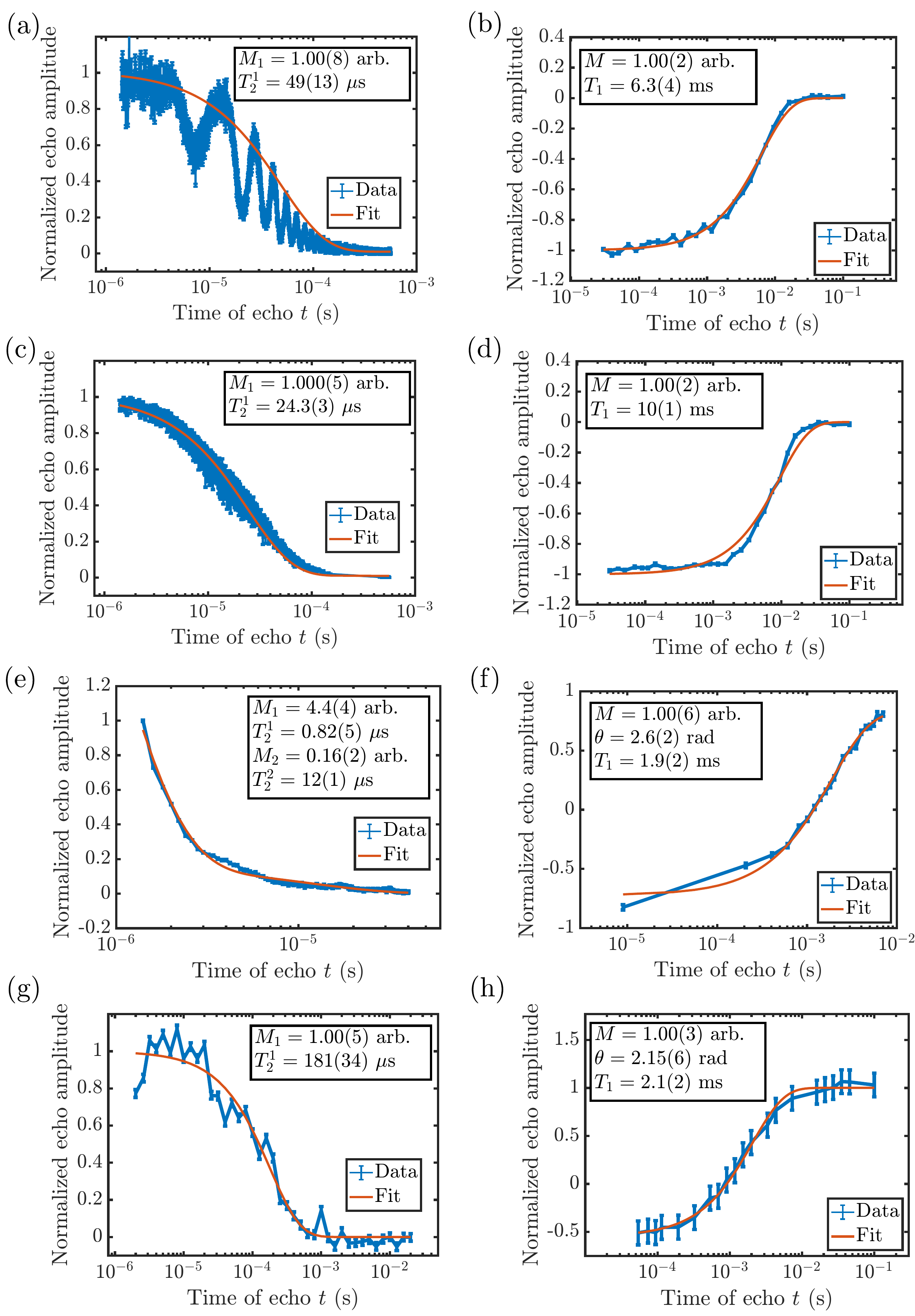}
\caption{ $T_1$ and $T_2$ data. \\
(a) Hahn echo, Sample A, NV center, 13 mT. \\
(b) Inversion depolarization, Sample A, NV center, 13 mT. \\
(c) Hahn echo, Sample A, NV  center, 190 mT. \\
(d) Inversion depolarization, Sample A, NV center, 190 mT. \\
(e) Hahn echo, Sample A, P1 center, 89 mT. \\
(f) Inversion recovery, Sample A, P1 center, 89 mT. \\
(g) Hahn echo, Sample B, P1 center, 89 mT. \\
(h) Inversion recovery, Sample B, P1 center, 89 mT. 
}
\label{fig:T1T2Exps}
\end{figure}

There are two diamond samples used in this work. Sample A is a high pressure, high temperature (HPHT) Type Ib diamond from Element 6 with a stone-type cut with approximate dimensions $4.2 \times 5.7 \times 1.3$ mm$^3$. Sample B is a chemical vapor deposition (CVD) diamond plate also from Element 6 with approximate dimensions $3.25 \times 3.17 \times 0.28$ mm$^3$. Here we discuss the $T_1$ and $T_2$ experiments that we performed to characterize these samples. Data are shown in Figure \ref{fig:T1T2Exps}, and resulting fit values are listed in Table \ref{Tbl:T1T2FitVals}.

Sample A has been processed for the generation of NV centers. The sample received a 1 MeV electron irradiation dose of $2\times10^{18}$/cm$^2$ per side followed by 2 hours of annealing at 800$^\circ$ C performed by US Diamond Technologies. This sample is reported by Element 6 to have a nitrogen concentration of [N] $\lesssim$ 200 ppm.
van Wyk \textit{et al.} \cite{wyk_dependences_1997} measured the Hahn echo $T_2$ of P1 centers to depend on nitrogen concentration as $T_2([\textrm{N}])= 71 \ \mu\textrm{s} \cdot \textrm{ppm} / [\textrm{N}]$. We have performed Hahn echo experiments on the P1 center (central ($m_I=0$) $^{14}$N hyperfine manifold) and fit the decay to a bi-exponential of the form.
\begin{equation}
    L_\textrm{HE}(t)=M_1 e^{-t/T_2^1}+M_2e^{-t/T_2^2}.
    \label{Eqn:T2BiExp}
\end{equation}
A bi-exponential was found to give the best fit for early and late decay times, which is indicative of the bulk sample containing a heterogeneous mixture of nitrogen concentrations with some more dense clustering. Recent studies suggest clustering of P1 centers is prevalent in HPHT Type Ib samples \cite{li_determination_2021, shimon_large_2022, bussandri_p1_2023, nir-arad_nitrogen_2023}. Measurement of the P1 Hahn echo decay curve is shown in Figure \ref{fig:T1T2Exps}(e). The dominant relaxation term in the fit, $T_2^1$, would imply a concentration of $[\textrm{N}]=87(8)$ ppm, with the reported uncertainty based on the 95\% confidence bound of the $T_2^1$ fit. To estimate the concentration of NV centers in Sample A, we multiply the nitrogen concentration by the conversion efficiency of the NV synthesis process. The conversion efficiency, estimated at 2\%, was determined by integrating the respective P1 and NV spectral lines of a continuous wave EPR spectrum obtained on another HPHT diamond sample that was in the same NV synthesis batch with Sample A. This spectrum was performed at the National High Magnetic Field Laboratory in Tallahassee, FL. We estimate the NV concentration in Sample A to be 1.7(3) ppm.

\begin{table}[]
\renewcommand*{\arraystretch}{1.3}
\caption{\label{Tbl:T1T2FitVals} Fit parameters for $T_1$ and $T_2$ experiments performed on the samples in this work. Data are shown in Figure \ref{fig:T1T2Exps}. The concentration characterizations for the P1 centers are based on $T_2$ measurements and the van Wyk \cite{wyk_dependences_1997} characterization. The concentration characterization for the NV centers is based on the P1 concentration in Sample A and continuous wave EPR spectra obtained at the National High Magnetic Field Laboratory. Uncertainties are based on the uncertainty of Hahn echo $T_2$ fits. We emphasize that, due to the heterogeneity of nitrogen \cite{li_determination_2021,shimon_large_2022,bussandri_p1_2023, nir-arad_nitrogen_2023}, these concentrations are only rough averages over the whole sample. $T_1$ is in reference to Equations \ref{Eqn:T1IR} and \ref{Eqn:T1ID}. For the NV centers in Sample A and the P1 centers in Sample B, the parameters $M_1$ and $T_2^1$ refer to fitting the Hahn echo decay to an exponential $M_1e^{-t/T_2^1}$. For the P1 centers in Sample A, the fit parameters $M_1$, $T_2^1$, $M_2$, and $T_2^2$ refer to the fit with Equation \ref{Eqn:T2BiExp}.}
\begin{tabular}{|c||c|c|c|c|}
\hline
Sample              & A   & A  & A   & B  \\ \hline
Species             & NV  & NV & P1  & P1 \\ \hline
$B_0$ (mT)          & 190 & 13 & 89  & 89 \\ \hline
Conc. (ppm) & 1.7(3)   & 1.7(3)  & 87(8) & 0.39(9)  \\ \hline
$T_1$ (ms)          & 10(1) & 6.3(4) & 2.1(2)  & 1.9(2)   \\ \hline 
$M_1$ (arb.)            & 1.00(5) & 1.00(8) & 4.4(4) & 1.00(5)   \\ \hline
$T_2^1$ ($\mu$s)   & 24.3(3) & 49(13) & 0.82(5) & 181(34)   \\ \hline
$M_2$ (arb.)            & - & - & 0.16(2) &  -  \\ \hline
$T_2^2$ ($\mu$s)   & - & - & 12(1) & -  \\ \hline
\end{tabular}
\end{table}

We have also fit the Hahn echo decays of the NV center in the HPHT sample. These are the CPMG-1 decays extracted from the same datasets shown in Figure 2(a) and (b). These are adequately represented by a simple exponential decay $L_\textrm{HE}=Me^{-t/T_2}$. For the NV center at 13 mT, we apply the fit only to the extracted peaks of the hyperfine modulations. According to Bauch \textit{et al.} \cite{bauch_decoherence_2020} who performed a survey of ODMR experiments on NV center ensembles in diamonds of widely varying nitrogen concentration, the NV $T_2$ is given by $T_2 =  (160 \pm 12) \, \mu\textrm{s ppm} / [\textrm{N}] \textrm{ ppm } $. Based on this characterization, we would expect coherence times of only 0.8(2) $\mu$s for the NV center.  However, as noted earlier, this relationship may not work well in the presence of heterogeneous P1 distributions. 
Our longer $T_2$ values may be measured from centers that have lower local P1 concentrations. Results from a recent theoretical investigation \cite{park_decoherence_2022} suggest a $T_2$ that is closer to our measured value but is still lower by about a factor of 10.

Sample B is reported by Element 6 to have $\textrm{[N]} < 1$ ppm. Based on the van Wyk characterization \cite{wyk_dependences_1997}, the 180(30) $\mu$s $T_2$ value we measured indicates a concentration of [N] = 0.39(9) ppm.

To characterize $T_1$ in our samples we performed inversion recovery (IR) experiments: $\pi - \tau_1 - \pi/2 - \tau_2 - \pi - \tau_2 - \textrm{acq.}$, where $\tau_1$ is varied over a wide range of delay times, $\tau_2$ is fixed at a small value for efficient readout, and ``acq.'' indicates the acquisition of the echo signal. For the P1 characterization, we fit the echo amplitude data to
\begin{equation}
    S_\textrm{IR}(t) = M(1-2\sin^2(\theta/2))\exp\left[ -t/T_1 \right],
    \label{Eqn:T1IR}
\end{equation}
where $t = \tau_1 + 2\tau_2,$ $M$ is a normalization term, and $\theta$ is the inversion flip angle whose deviation from $\pi$ indicates imperfect inversion of the spin ensemble. For the NV centers, the same pulse sequence is used, however it is preceded by the laser hyperpolarization pulse. The NV centers' thermal equilibrium signal is less than 1\% of the magnitude of the hyperpolarized signal. Accordingly, we refer to the NV $T_1$ experiment as inversion depolarization (ID), and fit the decay to
\begin{equation}
    S_\textrm{ID}(t) = M\exp\left[-t/T_1 \right].
    \label{Eqn:T1ID}
\end{equation}

For the experiments on Sample A, the diamond stone was oriented with $[111] \parallel \vec{B}_0$, whereas with the P1 experiments on Sample B, the diamond plate was oriented with $[100] \parallel \vec{B}_0$. The $T_1$ and NV concentration values are in fair agreement with those reported in \cite{jarmola_longitudinal_2015}, albeit at higher magnetic fields.

\section{Echo-detected field-swept spectra}
\label{sisec:EDFS}

\begin{figure*}[ht]
\includegraphics[width=\textwidth]{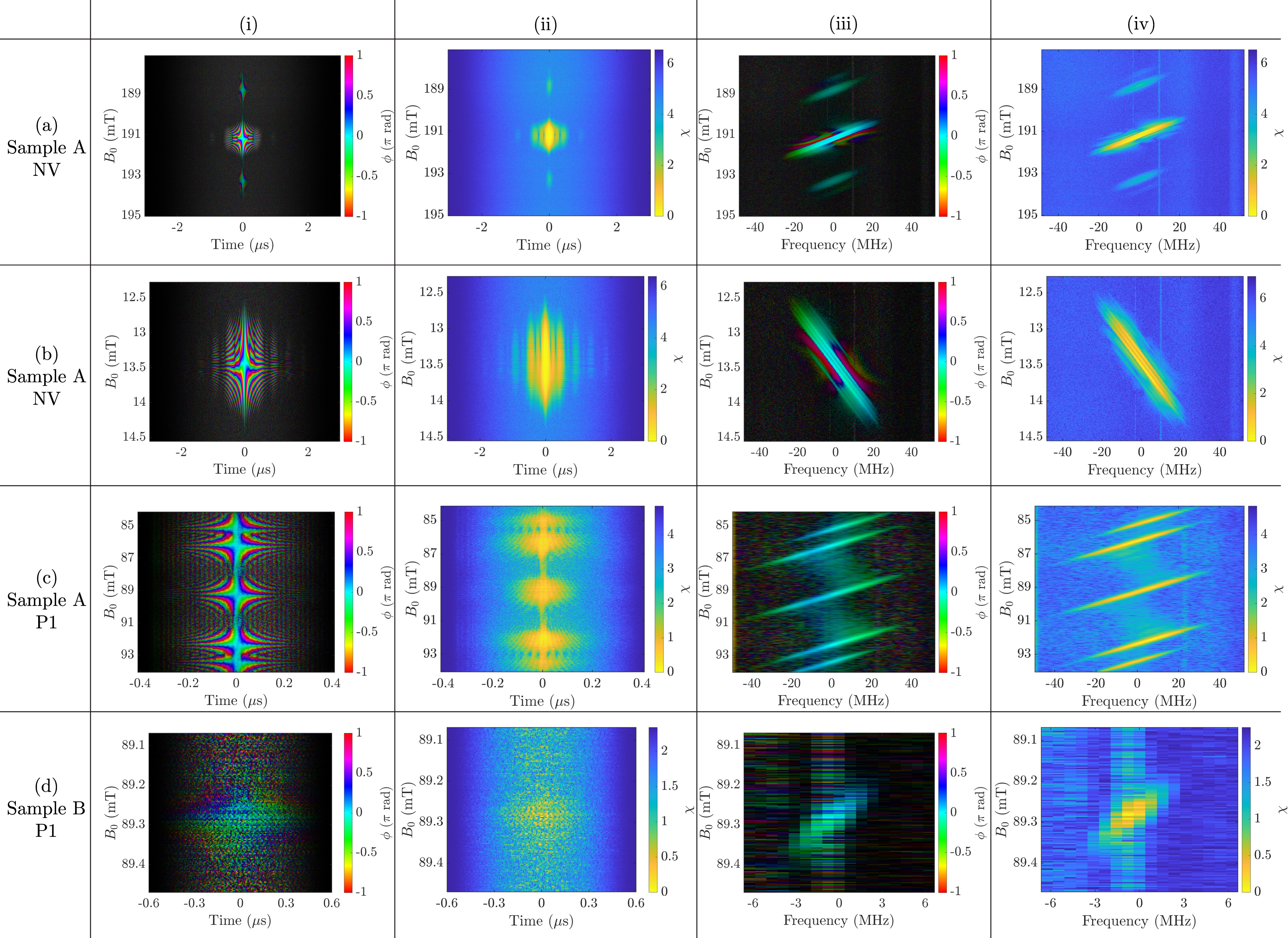}
\caption{\label{fig:E_ImageSpecs} Echo-detected field sweeps. Each column shows a different way of visualizing the data for the experiment (sample and spin species) specified by the label at left. The samples are described in Section \ref{sisec:Samples}. Column (i) shows the time trace of the echo signal with phase indicated by the color and log of the magnitude indicated by the brightness. Column (ii) shows the negative log of the signal magnitude, i.e. $\chi$, indicated by the color. Columns (iii) and (iv) are likewise visualizations of the phase and magnitude, but they show the Fourier transforms of each echo time trace. The vertical axes indicate the magnetic field.}
\end{figure*}

\begin{figure*}[ht]
\includegraphics[width=\textwidth]{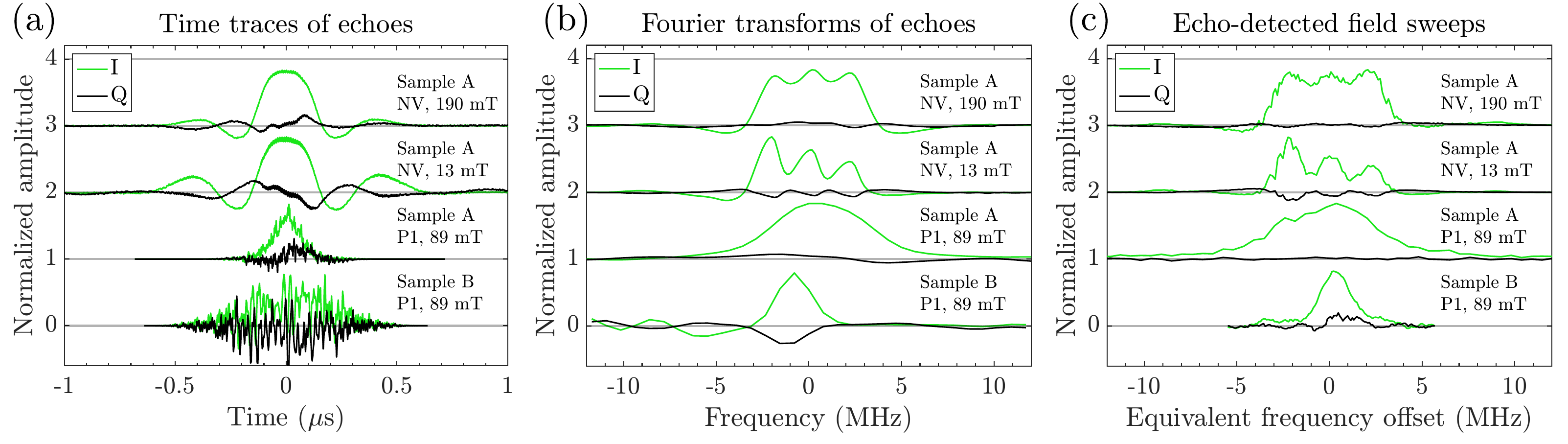}
\caption{\label{fig:ESpectra} (a) Example time traces of echoes extracted from the middles of the echo-detected field sweep spectra shown in Figure \ref{fig:E_ImageSpecs}. The real and imaginary component of each echo is shown, phase-corrected so that the real component has maximum amplitude. (b) Fourier transforms of the echo time traces in (a). (c) Echo-detected field sweep spectra. These are plots of the real and imaginary components along the line Frequency = 0 of the image plots shown in Column (iii) of Figure \ref{fig:E_ImageSpecs}. The Equivalent frequency offset is calculated by multiplying the magnetic field displacement by the gyromagnetic ratio for the electron.}
\end{figure*}

Examples of echo-detected field-swept spectra are shown in Figures \ref{fig:E_ImageSpecs} and \ref{fig:ESpectra}. Time traces of echoes obtained at a range of different magnetic fields are shown in Figure \ref{fig:E_ImageSpecs} Columns (i) and (ii) which show the phase and log magnitude of the signal respectively. In the time traces, Time = 0 is the middle of the echo. A Blackman window is applied to the time traces to suppress noise at the edges of the acquisition buffer. Columns (iii) and (iv) show the phase and log magnitude of the Fourier transforms of each time trace.

Figure \ref{fig:E_ImageSpecs} Row (a) shows a wide field sweep over the NV center around 190 mT. The CPMG experiments, referred to throughout this work as the 190 mT experiments, were performed by setting the magnetic field to the middle of the main spectral line, which occurs around 191.2 mT. The outer peaks near 189 mT and 194 mT are due to hyperfine coupling between NV centers and nearby \car nuclei. The primary NV spectral line can be seen in detail in the narrower field swept range of 13 mT data of Row (b). The NV echo time traces show a ripple, which, by examining the Fourier transforms, is understood to be an interference pattern caused by the excitation of the three spectrally separated lines arising due to the hyperfine interaction with the host $^{14}$N nuclei, clearly visible in the frequency trace shown in Figure \ref{fig:ESpectra}(b).

Figure \ref{fig:E_ImageSpecs} Row (c) shows the Sample A P1 spectrum. The four outermost spectral lines are those arising due to the hyperfine interaction with the host $^{14}$N nucleus. Due to the Jahn-Teller distortion \cite{smith_electron-spin_1959, bauch_decoherence_2020} and Sample A being in the $\vec{B}_0 \parallel [111]$ orientation, the ratio of the magnitudes of the inner to outer hyperfine-shifted lines is 3:1. Figure \ref{fig:E_ImageSpecs} Row (d) shows the Sample B P1 spectrum over a narrower field range than the Sample A P1 spectrum.

In Figure \ref{fig:ESpectra}(a) and (b), the example time traces and corresponding Fourier transforms are presented for signals acquired at a magnetic field in the middle of the primary peaks of the respective spectra. Figure \ref{fig:ESpectra}(c) shows the spectrum obtained by taking the dc value of the Fourier transformed time-traces at each magnetic field. The comparison of Figures \ref{fig:ESpectra}(b) and (c) reveals how the spectral line is embedded in the frequency content of the echo signal.

\section{Spectrometer and data collection}
\label{sisec:Spectrometer}

In order to explore the coherence properties of bulk electron spin systems we have constructed a pulse electron paramagnetic resonance (pEPR) spectrometer with an operating frequency of 2.5 GHz. It is worth emphasizing that whereas most other NV diamond experiments use optically detected magnetic resonance (ODMR), we use microwave detection, which is conventional in pEPR. We measure a microwave signal that gives the magnitude and phase of the transverse component of the net magnetization of the sample \cite{slichter_principles_1990}. This is achieved with the aid of a loop-gap resonator \cite{hardy_splitring_1981, froncisz_loop-gap_1982}, which houses the diamond sample as depicted in Figure 1(c). The loop-gap resonator is tuned by inserting a shard of dielectric (sapphire) into the gap. The coupling is adjusted by positioning an antenna a few mm above the top of the resonator \cite{joshi_adjustable_2020}. We are able to adjust the $Q$-factor of the resonator over a wide range this way. For the P1 experiments with Sample B, $Q = 2000$, and for the NV and P1 experiments with Sample A, $Q = 190$. The characteristic ring-down time is $2Q/\omega_0$, which is approximately 250 ns and 24 ns for the  respective experiments. The large $Q$ was used for the Sample B experiments to enhance sensitivity. The finite ring-down effects for these different $Q$ values are discussed in Appendix B and Section \ref{sisec:NumMeth}.

\begin{figure}[!hbt]
\includegraphics[width=0.8\linewidth]{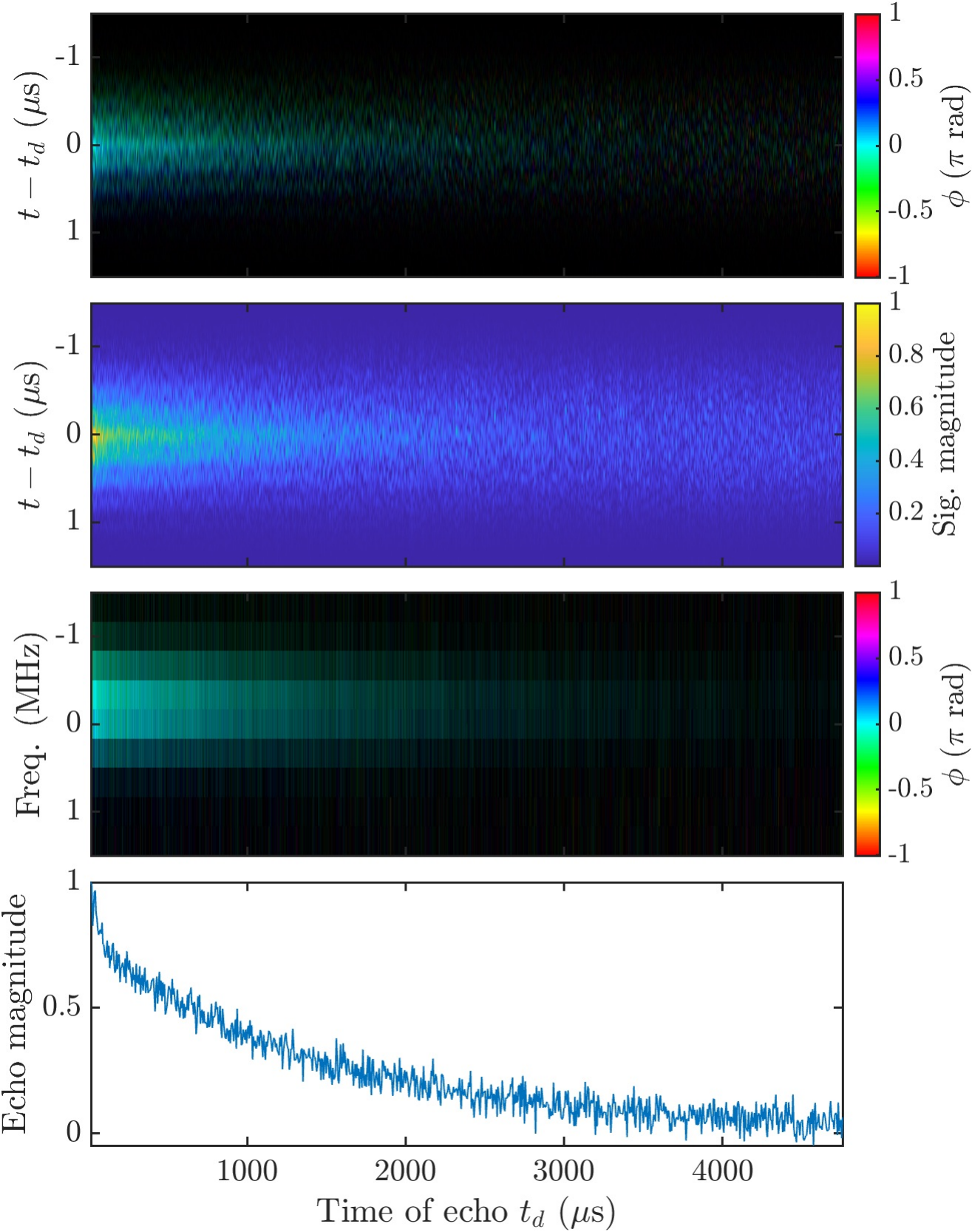}
\caption{\label{fig:LowNCPmg} Example recordings of a CPMG experiment with $\tau = 5.4 \ \mu$s recorded in the 89 mT data on the P1 centers of Sample B. The time of the echo $t$ is indicated in the bottom plot and is common to all 4 plots. The top two plots show the time traces of the echo, with 0 on the vertical axis indicating the center of the echo. The third plot shows the Fourier transform of each echo time trace. For the first and third plot, the color indicates signal phase, and the brightness indicates signal magnitude. The signal magnitude for the time traces is also indicated in the second plot. The echo magnitude, shown in the fourth plot, is the integral of the Fourier transformed time trace.}
\end{figure}

\begin{figure*}[t]
\includegraphics[width=\textwidth]{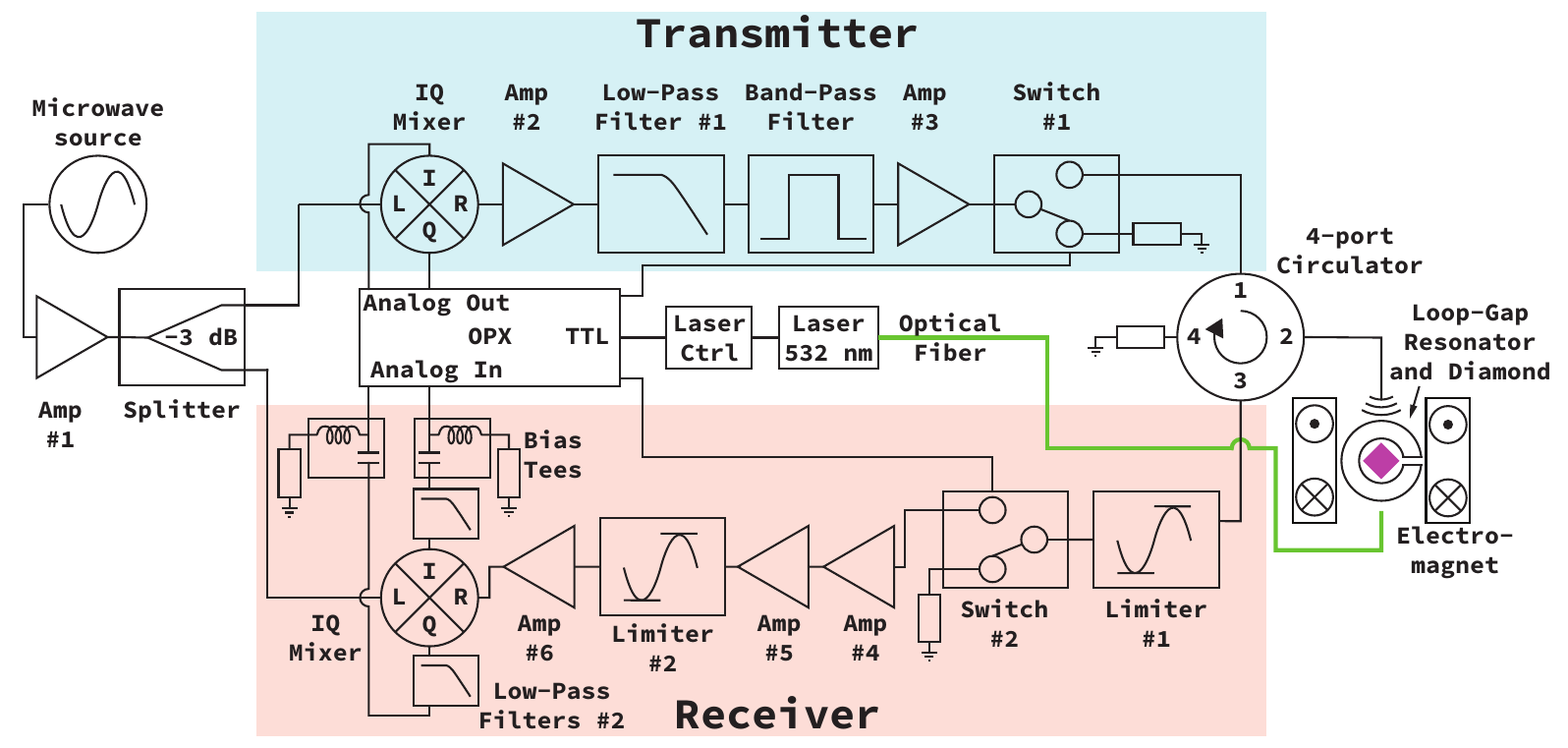}
\caption{\label{fig:MwSchem} Microwave circuit schematic of the pEPR spectrometer. See text for details.}
\end{figure*}

The schematic of our pEPR spectrometer is shown in Figure \ref{fig:MwSchem}. This is the instrument that was used to perform the experiments on Sample A, the HPHT diamond. Here we provide details of the spectrometer's construction and performance. The experiments on the P1 centers in Sample B, the CVD diamond, were performed on an earlier version of the instrument with a few differences mentioned later.

The microwave source is a Windfreak SynthUSB3, which outputs a continuous local oscillator (LO) signal at 2.8 GHz and +8.00 dBm. The LO signal is amplified (Amp \#1, Mini-Circuits ZX60-6013E-S+) and then split (Splitter, Mini-Circuits ZAPD-30-S+) between the Transmitter and Receiver arms of the spectrometer. In the Transmitter, the LO is mixed (IQ Mixer, Marki IQ1545LMP) with the analog outputs of the OPX (Quantum Machines) in a single-sideband mixing configuration. The OPX outputs pulses defined on a 300 MHz carrier to the I and Q ports of the mixer. The lower-frequency sideband coming out of the mixer contains the pulse on a 2.5 GHz carrier. The pulse is amplified (Amp \#2 same as \#1) then sent through Low-Pass Filter \#1 (Mini-Circuits VLF-2250+) and the Band-Pass Filter (Anatech Electronics, AB2593B765). Although the single-sideband mixing configuration with corrective biasing should eliminate unwanted sidebands, LO leakage, and harmonics, the extra filtering ensures the attenuation of any signal other than the desired pulse. The pulse signal is then amplified by the power amplifier (Amp \#3, Mini-Circuits, ZHL-2425-250X+), which we restrict to a maximum power of +49 dBm. Switch \#1 (DBwave Technologies, PASR0200500600B) transmits the pulse when the spectrometer is in transmit mode, and routes amplifier noise into a 50 $\Omega$ termination at all other times. The switching is controlled by an OPX TTL line. The pulse signal enters port 1 of the 4-port circulator (Ditom Microwave, D4C2327) and is routed out port 2 to the loop-gap resonator. The echo signal returns to port 2 and is routed to the receiver. Port 4 of the circulator is capped with a 50 $\Omega$ termination to isolate the transmitter and damp out stray reflections. Limiter \#1 (Fairview Microwave, FMMT1021) is inserted at the start of the receiver chain to protect Switch \#2 (Mini-Circuits, ZFSWA2-63DR+) from the reflections of the high-power pulses. When receiving an echo signal, Switch \#2 routes the echo to the amplifier chain, and at other times it routes signals to a 40 dB, 50 W attenuator (Mini-Circuits BW-N40W50+) followed by a 50 $\Omega$ termination. After Switch \#2, the echo signal is picked up by two low-noise amplifiers in series: Amp \#4 (RF-Lambda, RLNA02G02G) followed by Amp \#5 (Mini-Circuits, ZQL-2700MLNW+).
Limiter \#2 (Mini-Circuits, ZFLM-252-1WL-S+) prevents any damage from pulse reflections that leak through Switch \#2. The echo signal is amplified once more (Amp \#6, same as Amps \#1 and \#2) before being downconverted from 2.5 GHz to 300 MHz by the receiver IQ mixer (same kind as transmitter IQ mixer). The I and Q outputs of the receiver IQ mixer are sent through Low-Pass Filters \#2 (Mini-Circuits, SLP-1650+) to remove the 5.3 GHz sideband. The filtered in-phase and quadrature components of the echo signal are passed through equal-length transmission cables to bias tees (Mini-Circuits, ZFTB-4R2G+) for AC coupled analog signal acquisition by the OPX. The OPX also controls the triggering of the laser controller (CivilLaser, LSR532H-1W-FC). The 532 nm laser light is transmitted to the sample via an optical fiber (Thorlabs, FG365UEC), the ferrule of which is mounted in the 3D-printed assembly as depicted in Figure 1(c).

The experiments on P1 centers in Sample B were performed on an earlier version of this spectrometer. In the earlier version, the power amplifier in place of Amp \#3 (Spectrian, 20 W 2.4 GHz) output 30 dBm, and in place of Switch \#1 were two Mini-Circuits ZFSWA2R-63DR+ switches. It did not include any laser optics. The earlier version did not have an OPX. Instead, a Tektronix AWG7052 was used for pulse signal generation on a 100 MHz carrier which was up-converted with a 2.4 GHz LO to 2.5 GHz. The AWG7052 was also used for triggering the switches and the digitizer (SP Devices ADQ214). A SpinCore PulseBlaster ESR-PRO-500-PCI was used to trigger the AWG7052.

A schematic of the CPMG sequence is shown in Figure 1(b). For experiments on NV centers, initialization into the $m=0$ state was achieved with 532 nm laser light. The initialization laser pulse is 10 ms in duration. 10 ms was found to be the laser pulse duration at which the NV signal magnitude saturates. The laser is controlled simply by triggering the laser controller. The controller has a maximum modulation rate of 30 kHz. Accordingly, we include a 100 $\mu$s delay after the end of the laser pulse before starting the microwave pulses.  It is worth noting that this initialization process is significantly different from ODMR NV experiments, which typically have laser initialization pulses and delays on the order of a few $\mu$s, achieved with acousto-optic modulators. The most significant reason for the difference is that we are sending unfocused light into an entire diamond sample, instead of focused light into a microscopic region, so the intensity of light in our sensitive volume, about 1 W/mm$^2$, is accordingly much lower than that in an ODMR experiment.

pEPR measurements are weak measurements in that the coupling between an electron spin and the resonator is weak enough that there is no back-action that could disrupt the state of the spin. This enables us to perform stroboscopic acquisitions of the CPMG echoes. For a single $\tau$ value, in a single shot, we capture every echo out to a maximum detection time of about 3 ms, at which point the echoes have largely faded into the noise. For each acquisition, the signal is demodulated and integrated in order to obtain the amplitude of the echo. Examples of the echo time traces and processing steps for a CPMG experiment on the P1 centers in Sample B are shown in Figure \ref{fig:LowNCPmg}. In this example, the recording buffer is 3 $\mu$s in length. We apply a Blackman window to the time trace before integration. For the CPMG experiments with Sample A, the acquisition buffer is 400 ns. The demodulation is performed onboard the OPX, and we record just a pair of values (I and Q), such that the echo magnitude is $\sqrt{\textrm{I}^2 + \textrm{Q}^2}$. After the microwave sequence is finished, the NV centers are re-initialized with the laser, and the sequence is repeated for averaging. For the NV center experiments we perform 128 averages.

For the P1 measurements, there is no direct hyperpolarization mechanism, so no laser light is shown on the sample. The P1 signal magnitude is proportional to the polarization of the P1 center in the applied $B_0$ field at thermal equilibrium. 

We wait 5 ms, $\approx 2.5 \times T_1$, in between shots. For the P1 centers in Sample A (B), 16,384 (65,536) averages were performed.

The minimum values of the interpulse delay $\tau$ that we can achieve are limited by a variety of factors. For the Sample B P1 CPMG experiments, the minimum $\tau$ presented in this work is 4 $\mu$s. At lower $\tau$ values, the 250 ns ring-down of the resonator begins to interfere with the echo signal. In the experiments on Sample A, the minimum $\tau$ value presented is 1.4 $\mu$s. The rise/fall times of the switches plus the finite pulse and recording durations put the minimum possible $\tau$ around 1 $\mu$s. Additionally, we find that Limiter \#1 saturates for $\tau$ of 1.4 $\mu$s after 3 ms of applying $\pi$ pulses. We restrict the $\tau$ and sequence duration to these values to avoid saturating the limiter.

We performed the NV CPMG experiments at $\tau$ values ranging from 1.4 $\mu$s to 560 $\mu$s. As explained in Appendix A, the harmonic analysis for the \car peak characterization is restricted to $\tau \leq 10 \ \mu$s. For the 190 mT experiment, $\tau$ was sampled linearly up to $\tau = 10 \ \mu$s, with a step of 8 ns. Above $\tau = 10 \ \mu$s, $\tau$ was sampled logarithmically for efficiency. For the 13 mT experiment, the linear to logarithmic sampling transition was at 6.8 $\mu$s. In executing an experiment, the $\tau$ values were selected in a random order. All 128 averages are acquired for the CPMG decay curve with one $\tau$ value before continuing to the next $\tau$ value. One acquisition loop iteration consists of the laser initialization (10.1 ms), the microwave pulse sequence (3 ms), and a hardware recovery delay (60 ms), such that acquiring all 128 averages for a given $\tau$ takes about 9.3 s. Between runs for different $\tau$ values, the data is saved, and the magnetic field control loop is executed, which typically takes 4.5 s. So, on average, it takes 13.8 s to acquire the data for a given $\tau$ and proceed to the next. For the 13 mT experiment, a total of 4356 $\tau$ values were sampled, taking about 16.7 hours. For the 190 mT experiment, a total of 6447 $\tau$ values were sampled, taking about 24.7 hours.

The Sample A P1 CPMG experiments linearly sampled $\tau$ from 1.4 $\mu$s to 19.96 $\mu$s, with a spacing of 80 ns. Experiments included a 500 $\mu$s microwave pulse sequence and a 70 ms measure delay to allow for hardware recovery. 16,384 repetitions were performed for each $\tau$ value. In total, the 233 experiments took 80 hours to complete, for an average of about 21 minutes for each $\tau$.

For the Sample B P1 CPMG experiments, $\tau$ was sampled linearly from 4 to 41.3 $\mu$s with a step of 100 ns. No laser was used. The microwave pulse sequences were applied for 5 ms, followed by a spin system recovery delay of 5 ms (about $2.6 \times T_1$). This sequence was repeated until the digitizer's memory limit was reached, at which point each individual time trace was transferred to the lab computer, and field correction was performed. This was repeated until the 16,384 averages of the CPMG decay were obtained for a given $\tau$ value and a given global phase of the microwave pulses. This was repeated for global phases of the control pulses of 0, $\pi/2$, $\pi$, and $3\pi/2$, and the signals were added with appropriate phase corrections for a total of 65,536 averages. This phase-cycled averaging eliminated phase-independent systematic noise. Due to the much higher number of averages needed and the overhead time of data transfer, this experiment took 31 days to complete with an average of 2 hours to acquire data for each of the 373 $\tau$ values.

\section{Discussion of 1/$\omega^{0.7-1.0}$ spectrum}

Observing a power spectrum of the form $1/\omega$ would happen if the varied P1 concentrations throughout the sample give rise to an effective distribution of Lorentzian power spectra with correlation times $\tau_c$ that can be approximated as $P(\tau_c) = P_0/\tau_c$ between some range of cutoff correlation times $\tau_1 < \tau_c < \tau_2$. Then the ensemble average power spectral density would be
\begin{eqnarray}
   \langle S(\omega) \rangle_\textrm{ens} &= 2 \Delta^2 P_0 \int_{\tau_1}^{\tau_2} \frac{d \tau_c}{1 + \omega^2 \tau_c^2} \\
   &= \frac{2 \Delta^2 P_0}{\omega}(\arctan(\tau_2 \omega) - \arctan(\tau_1 \omega)),
\end{eqnarray}
which takes the form of $1/\omega$ for $1/\tau_2 \ll \omega \ll 1/\tau_1$. 
Typically, $T_2\propto[\textrm{N}]^{-1}$, as has been shown with simulations \cite{wang_spin_2013} and experiments \cite{wyk_dependences_1997, bauch_decoherence_2020}. Considering also that $T_2 \propto \tau_c$, this suggests that the $1/\omega$ noise could arise for a heterogeneous population of P1 concentrations ([N]) with an approximate distribution throughout the sample of $P([\textrm{N}]) \propto [\textrm{N}]$ over some range of [N] corresponding to $\tau_1$ and $\tau_2$. Further research would be needed to confirm or rule out either of these explanations.  

\begin{figure*}[htb]
\includegraphics[width=\linewidth]{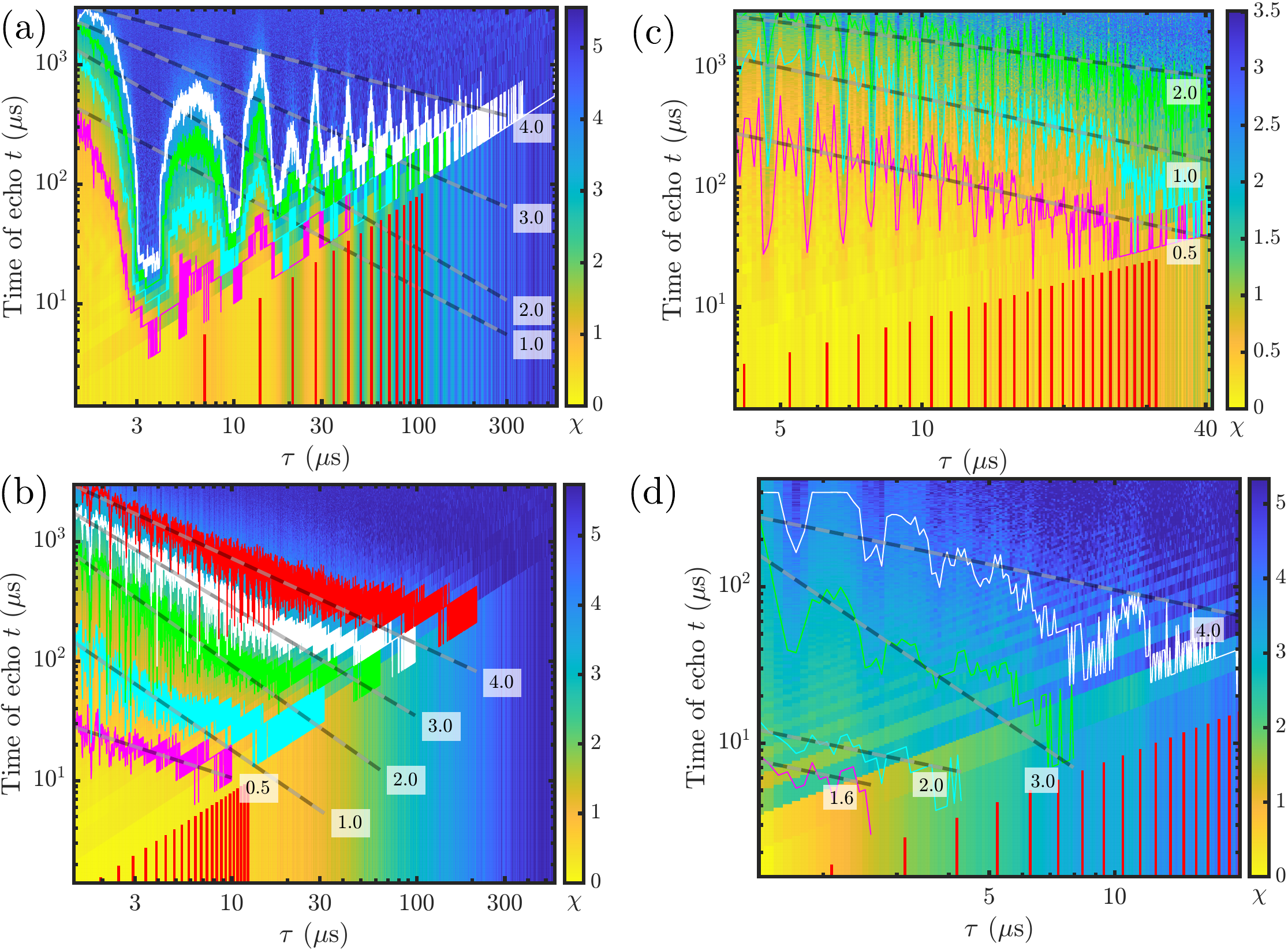}
\caption{\label{fig:2_v2Contours} The same image plots as shown in Fig. 2 with $\chi$ contour data resulting from the contour tracing analysis step added. The contour data shown are used for fitting the straight line power law fits as described in Sec. IV.A.}
\end{figure*}

\section{Properties of the \car peak measurements}
\label{sisec:AmplitudeCar13}

Discussion of the differences between the \car power spectrum properties measured with the Sample A P1 centers vs. the other experiments are covered in the main text in Section IV C. Here we discuss the more subtle differences in the peak measurement among the lower concentration species (Sample A NV centers and Sample B P1 centers).

The fits for the \car peak widths $\sigma$ in Figure 5(d) show consistency between the Sample B P1  measurement at 89 mT and the NV measurement at 190 mT. However, the width of the NV peak at 13 mT is approximately 70\% greater than the width of the other two measurements. The reason for this is believed to be that at 13 mT, the \car nuclear spins are in a regime where the NV hyperfine and Zeeman terms are equal for a \car nuclear spin \cite{reinhard_tuning_2012}. Reinhard \textit{et al.} \cite{reinhard_tuning_2012} define a critical radius around an NV center, where \car spins outside this radius are quantized along the direction of the external Zeeman field  while \car spins inside this radius are quantized along the local hyperfine axis. We estimate that the number of \car nuclei within the critical radius is $N_c(B_0) = 0.08 \textrm{ mT}/B_0$, which is 6 at 13 mT and 0.4 at 190 mT. The broader peak observed at 13 mT is due to the contributions of these hyperfine-shifted spins.

Figure 5(e) shows the percentage error of the \car frequency identified by the Gaussian fitting of the spectral line. We calculate the expected (``literature'') value as $\omega_\textrm{lit.} = B_0 \gamma_\textrm{13C}$, and we calculate $B_0$ as $|(D \pm \omega_\textrm{spec})/\gamma_e|$ with $\pm$ being $-$ for 13 mT and $+$ for 190 mT. The fractional uncertainty in the frequency measurement is $\delta D/D + \delta \omega_\textrm{spec}/\omega_\textrm{spec}$, which depends on the temperature variations in our lab and matching the resonant frequency and magnetic field. 
These fractional uncertainties of the \car frequencies are represented by the gray bands in the background of Figure 5(e). At all fields, the measured value is accurate within the uncertainty of the calculation. 

Based on the temperature-dependent studies of the NV resonance spectrum in Reference \cite{acosta_temperature_2010} and the temperature fluctuations in our laboratory, we use a value of $D=2.866(10)$ GHz for the 13 mT measurement and $D=2.865(10)$ GHz for the 190 mT measurement, which was slightly warmer due to heat from the magnet coils at the higher field. $D=0$ for the P1 center. The uncertainty in the effective spectrometer frequency depends on the centering and symmetry of the spectral line which forms the echo signal. Example echoes obtained in echo-detected field sweep spectra are shown in Figure \ref{fig:ESpectra}. 
We used $\omega_\textrm{spec} = 2.490(3) \times 2 \pi$ rad GHz in the NV experiments and $\omega_\textrm{spec}=2.500(3) \times 2 \pi$ rad GHz in the P1 Sample B experiment. For the P1 center, the fractional uncertainty of \car Larmor frequency is 0.0012, and for the NV centers it is 0.005.
These uncertainties are represented by the gray bands in the background of Figure 5(e). At all fields, the measured value is accurate within the uncertainty of the calculation. 
At 13 mT and 89 mT, the calculated uncertainties are comparable to the 95\% confidence intervals of the fits. At 190 mT, the frequency measurement has notably higher precision, demonstrating the utility of the harmonic analysis for frequency measurements at intermediate magnetic fields.

Characterizing the amplitude of the \car spectral peak is less straightforward than characterizing its frequency position or width, as estimates depend on the range of sub-harmonics used \cite{hernandez-gomez_noise_2018}.   Additionally, measuring sharp peaks in the power spectrum is significantly affected by flip angle (over/under rotation) errors of the CPMG $\pi$ pulse, which cause an inversion in the middle of the spectral peak \cite{lang_quantum_2019} and can shift the apparent frequency. It is also significantly affected by finite pulse widths, which effectively reduce the amplitudes of the peaks of the higher harmonics in the filter function. We have taken these considerations into account and explored their effects numerically and experimentally in Section \ref{sisec:NumMeth}.

\begin{figure}[ht]
\includegraphics[width=0.6\linewidth]{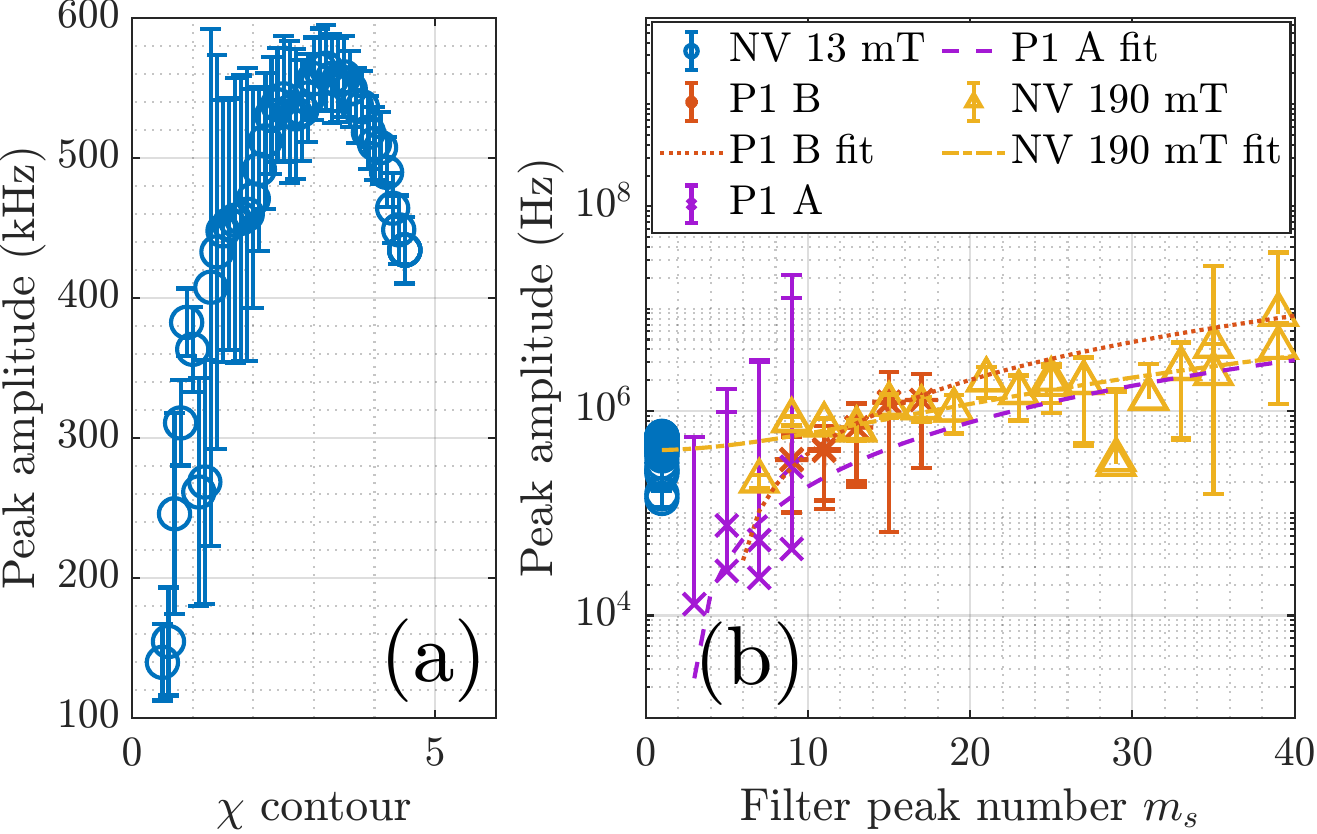}
\caption{\label{fig:C_PkOrds} (a) The \car peak amplitude at 13 mT measured by fitting $S(\omega_1)$ for different $\chi$ contours to a Gaussian of the form $Ae^{-(\omega-\omega_P)^2/(2\sigma^2)}$ with $A, \omega_P,$ and $\sigma$ as fit parameters. (b) Peak amplitudes from all three main data sets versus the filter harmonic $m_s$ that was used to scan the peak.}
\end{figure}

Consider the reconstructed $\chi=3.5$ fundamental spectrum of the 13 mT experiment in Figure 4(c). To reconstruct the fundamental spectrum, we used the Gaussian fit parameters from the primary peak at 140 kHz. The reconstructed subharmonic at 48 kHz appears to have only about half the amplitude of the subharmonic obtained directly from the experimental data. The subharmonics of the 89 mT and 190 mT data in Figures 4(a), (b) and (c) show this trend as well. The size of the peak as measured with different harmonics is shown in Figure \ref{fig:C_PkOrds}. We emphasize that the spectra obtained from the harmonic analysis shown in Figures 5(a), (b) and (c) represent the average effective \car peak over the valid range of inter-pulse delays $\tau$ and echo numbers $N$, as explained in Appendix A. Some of the results on a single NV center in Reference \cite{hernandez-gomez_noise_2018} also seem to suggest that the measurements of the \car peak with higher harmonics of the filter function return a larger value for the amplitude than measurements with lower harmonics.

The challenge with measuring the \car peak amplitude is that in the 13 mT data, the \car interaction is so strong that it causes complete decoherence at low $N$, whereas in the 89 mT and 190 mT data, the peak seems to be greater in magnitude when measured with larger $\tau$ (i.e., when a larger filter harmonic $m_s$ overlaps with the peak).

The same $\tau$ and $t$ limits that constrain the harmonic analysis (red dashed rectangles in Figures 8(a), 9(a) and 10(a)) apply to measuring the \car peak in the 13 mT data as well: the width of the filter peak should be significantly less than the width of the \car peak, which is about 10 to 20 kHz. Approximating the filter peak as a Gaussian gives $\sigma=\sqrt{2 \pi}/t$. Accordingly, to make use of the $\delta$-function approximation of the filter for the \car peak measurement at 13 mT, detection times of $t \gg \sqrt{2\pi}/(10 \textrm{ kHz})\approx 25 \ \mu$s ought to be used. It is clear however from Figure 2(a) that $\chi$ is nearing the noise threshold ($\chi \gtrsim 4$) at such detection times in the middle of the first coherence dip ($\tau \approx 3.5 \ \mu$s). In Figures 4(c) and 5(d), (e) and (f) we have presented and analyzed the $\chi=3.5$ contour for the 13 mT data since it is mostly observed at detection times later than 25 $\mu$s. 

Figure \ref{fig:C_PkOrds}(a) shows the amplitudes of the \car peak obtained by applying Gaussian fits to the different $\chi$ contours of the 13 mT data. For fits of $\chi<3.5$, the detection times are early, and there is considerable uncertainty due to the roughness of the $\chi$ contour at low $N$ (as in Figure 4(a) with $\chi=1$). For fits of $\chi>3.5$ the signal-to-noise (roughly $\log(5)/\log(\chi)$) becomes too small. Accordingly, the \car peak amplitude around $\chi=3.5$ is likely the most accurate measurement from the 13 mT data.

The \car peak amplitudes presented in Figure 5(f) for the 89 mT and 190 mT data represent the average over the sensing areas (red dashed rectangles of Figures 8(a), 9(a), and 10(a)). These average \car amplitudes are more heavily weighted toward the measurements with lower harmonics $m_s$, which are obtained with the shorter $\tau$ values. This is because there is a higher density of acquisitions at shorter $\tau$ values. Figures 8(d), 9(d), and 10(d) show the net contributions of the data to forming the spectrum from the harmonic analysis. Figures 8(g), 9(g), and 10(g) show the amplitude of the \car peak as obtained with each available scan harmonic $m_s$. There is an apparent increase of the \car peak amplitude as measured with increasing harmonic number $m_s$. However, this is likely largely due to the compounded uncertainty in the harmonic analysis procedure. Based on Equation 21, and approximating $\delta \chi \approx \delta \chi_B$, the uncertainty for the sharp peak, $S_P$, measurement is
\begin{equation}
    \delta S_P \approx \frac{m_s^2 \pi^2}{4 N \tau} \sqrt{2} \, \delta \chi_B.
\end{equation}
Due to the $m_s^2$ term, $S_P$ tends to grow quadratically with $m_s$. We have shown the peak amplitude fits as a function of the harmonic $m_s$ used to measure the peak for all three main data sets in Figure \ref{fig:C_PkOrds}(b). To show that the peaks in $S_P$ grow quadratically with $m_s$ due to uncertainty, we have fit the 89 mT and 190 mT peaks as $S_P^\textrm{(peak)}(m_s) = p_2 m_s^2 + p_0$, where $p_n$ is the coefficient of the $n^\textrm{th}$ order polynomial term. In the limit that $\delta S_P = 0$, our model predicts that $p_2 = 0$, and $S_P^\textrm{(peak)}=p_0$. That is, in a classical Gaussian noise spectrum, the measured amplitude of the peak should not depend on which filter harmonic $m_s$ is used to measure it (taking into account the appropriate weighting factors). These fits are depicted in Figures 8(g), 9(g), 10(g), and \ref{fig:C_PkOrds}(b) and show approximate agreement with the measurements. For the Sample B P1 data we obtain $p_2=5.4 \pm 1.4$ kHz and $p_0= -160 \pm 240$ kHz. For the Sample A P1 data, we obtain $p_2 = 2 \pm 3$ kHz and $p_0 = 0 \pm 200$ kHz. For the 190 mT NV data we obtain $p_2 = 1.9 \pm 1.1$ kHz, and $p_0 = 400 \pm 800$ kHz.

The difficulties of measuring the \car peak amplitude are apparent. We can place a lower bound on the peak amplitude of about 500 kHz. Likely, the most accurate Gaussian approximation of the peak is slightly higher than this, given by the lowest $m_s$ measurements from the 89 mT and 190 mT data. In this sense, the values reported in Figure 5(f) are the most accurate estimates. It is possible that the magnitude of the \car noise varies with magnetic field, but our measurements do not show that it is substantial.

\section{Numerical methods}
\label{sisec:NumMeth}
\subsection{Filter theory}
\label{sisec:FilterTheory}

The harmonic analysis method explained in this work relies on the high-frequency content of the CPMG filter function to perform spectroscopy. In the time domain, this high-frequency content is associated with sharp, instantaneous transitions between $+1$ and $-1$ in the modulation function. In real experiments, finite pulses smooth these transitions, creating a gradual change between $+1$ and $-1$. This smoothing in the time domain corresponds to a reduction of the high-frequency content of the filter. The modulation function also takes on a perpendicular component during the transition. This transverse component of the modulation function is also added to by flip angle errors. Here we explain how we numerically obtain the leading order correction to the decoherence calculation. We examine the case for CPMG with finite pulses and flip angle errors. For a generalized treatment of noise spectroscopy with multi-axis control, see Reference \cite{green_arbitrary_2013}.

In CPMG the initial density operator, following the $(\pi/2)_x$ pulse, is $\rho_0 = \frac{\mathbbm{1}}{2} - \epsilon S_y$. The control Hamiltonian for CPMG is $H_c(t) = \omega_1(t) S_y,$ where $\omega_1(t)$ is the amplitude of the control. The control unitary is
\begin{eqnarray}
    U_c(t) &= \exp\left[ -i S_y \int_0^t dt' \omega_1(t') \right] \\
    &= \exp\left[ -i S_y \Phi(t) \right],
\end{eqnarray}
where $\Phi(t)$ is the cumulative phase induced by the control pulses. In the interaction picture defined by the Zeeman and environment Hamiltonians, the effective system-environment Hamiltonian $H^{(E)}_{SE} = -\gamma_e b_z(t)S_z$. In the toggling frame, which is the interaction picture defined by the control pulses, the effective system-environment Hamiltonian becomes $\tilde{H}_{SE}(t) = U_c^\dag(t)H^{(E)}_{SE}U_c(t).$
The multi-axis filter function is $f_\mu(t) = 2 \textrm{Tr}\{S_\mu U_c^\dag(t) S_z U_c(t)\},$ which, in the case of CPMG with exclusively $y$-pulses, simplifies to 
\begin{equation}
\label{Eqn:ModFuncFromPhi}
\vec{f}(t)=(-\sin(\Phi(t)), 0, \cos(\Phi(t))).
\end{equation}
We can then express the system-environment Hamiltonian as $\tilde{H}_{SE}(t) = -\gamma_e b_z(t)(-\sin(\Phi(t)) S_x + \cos(\Phi(t))S_z).$ To estimate the decoherence due to this interaction between initialization and detection time ($t$), we approximate the Hamiltonian as a time-independent Hamiltonian using the Magnus expansion. The leading order term of the Magnus expansion is
\begin{eqnarray}
    \tilde{H}_{SE}^{(0)}(t) &= \frac{1}{t}\int_0^{t} \tilde{H}_{SE}(t') \, dt' \\
    &=-\frac{1}{t} \bigg( \int_0^{t} \gamma_e b_z(t')f_x(t')S_x \, dt' 
    + \int_0^{t} \gamma_e b_z(t')f_z(t')S_z \, dt' \bigg) \\
    &= \frac{1}{t} (\phi_x S_x + \phi_z S_z),
\end{eqnarray}
where $\phi_\mu$ is the net phase accumulated about the axis $\mu$ due to the noise. Higher-order terms of the Magnus expansion contain nested commutators of the form $[\tilde{H}_{SE}(t_1),\tilde{H}_{SE}(t_2)]$. In the case of perfect, instantaneous pulses $f_x(t)=0,$ so these commutators are 0 and the leading order term is exact. With real pulses, the higher-order terms will contribute to additional decoherence; however, in the weak-noise limit, their effect is small enough to be ignored \cite{green_arbitrary_2013}.

With this, we calculate the signal as the zeroth order approximation of the normalized ensemble average value of $S_y$ at time $t$:
\begin{eqnarray}
    \langle s^{(0)}(t) \rangle_{\textrm{ens}}
    &= \left\langle \frac{\textrm{Tr}\{ e^{-i \tilde{H}_{SE}^{(0)}(t)t} \rho_0 e^{i \tilde{H}_{SE}^{(0)}(t)t} S_y \}}{\textrm{Tr}\{\rho_0 S_y\}} \right\rangle_\textrm{ens}  \\
    &=\langle \cos(\phi_x) \cos(\phi_z) \rangle_\textrm{ens} \\
    &= \int_{-\infty}^\infty d\phi_z \int_{-\infty}^\infty d\phi_x \cos(\phi_x) \cos(\phi_z) P(\phi_x,\phi_z),
\end{eqnarray}
where $P(\phi_x,\phi_z)$ is the distribution of the random variables $\phi_x$ and $\phi_z$. Assuming $b_z(t)$ is a Gaussian random variable, then so too are $\phi_x(t)$ and $\phi_z(t)$. We then obtain
\begin{equation}
    \langle s^{(0)}(t) \rangle_{\textrm{ens}} = e^{-\langle \phi_x^2 \rangle/2 -\langle \phi_z^2 \rangle/2}
\end{equation}
leading, via the convolution theorem, to the overlap integral
\begin{eqnarray}
    \chi &\equiv (\langle\phi^2_x\rangle + \langle\phi^2_z\rangle)/2 \\
    &= \frac{t^2}{2} \left[ \int_{-\infty}^\infty d\omega' S(\omega')(F_x(\omega') + F_z(\omega')) \right]
    \label{Eqn:ChiMultAxFilt}
\end{eqnarray}
where $S(\omega)$ is the power spectrum of $b_z(t)$, and the $\mu$-axis filter function is
\begin{equation}
\label{Eqn:FilterMultiAx}
    F_\mu(\omega) = \frac{1}{t^2} \left| \frac{1}{\sqrt{2\pi}} \int_{-\infty}^{\infty} f_\mu(t) e^{-i \omega t} \, dt \right|^2.
\end{equation}
We numerically compute $F$ by constructing $\Phi(t)$ as a finely sampled array for a real pulse sequence based on our physical system, and then take the square of the magnitude of the discrete Fourier transform.

\begin{figure*}[t!]
\includegraphics[width=\linewidth]{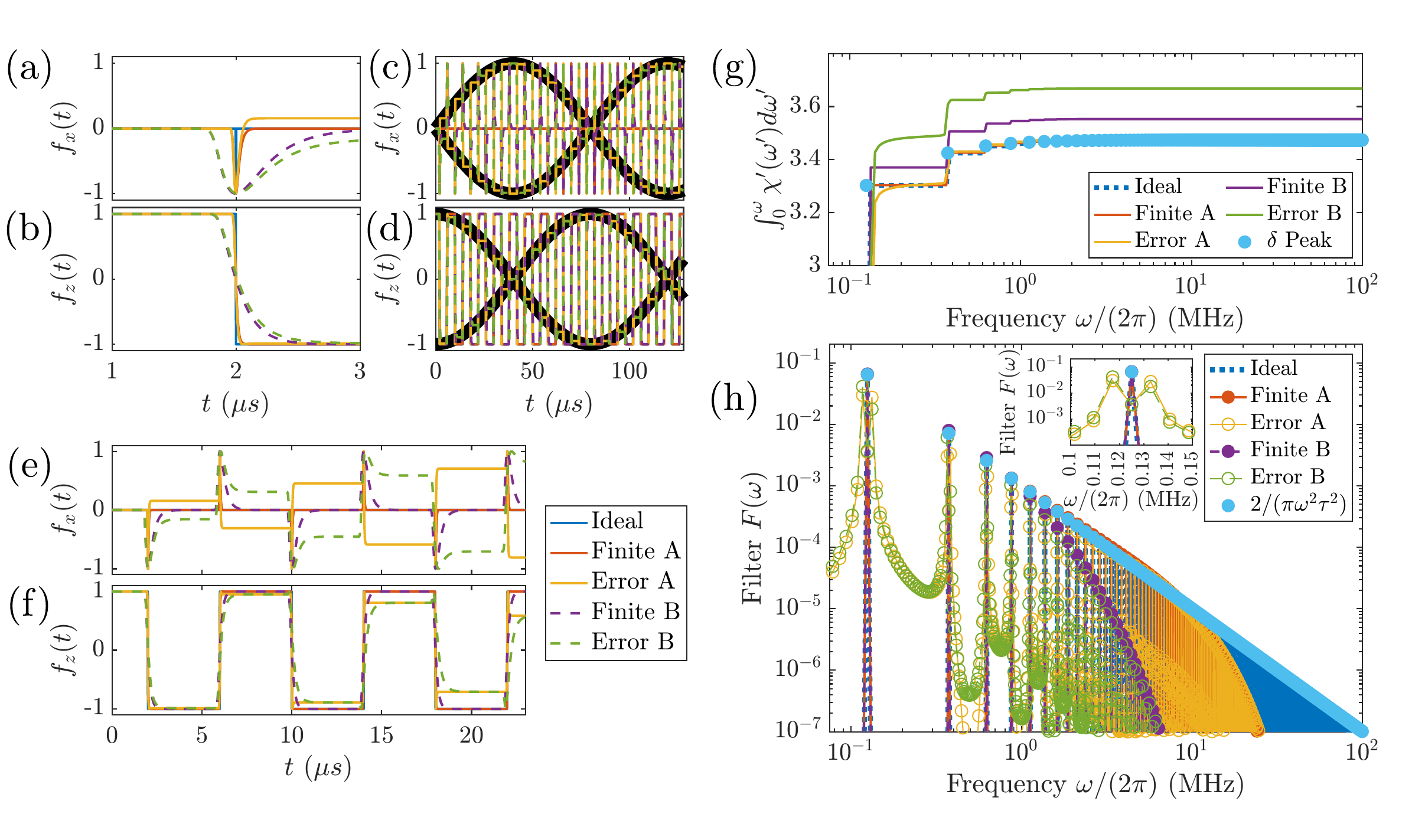}
\caption{\label{fig:FFinite}
Example calculations from the numerical investigations into finite pulse effects. (a), (c), and (e) show the $x$ components and (b), (d), and (f) show the $z$ components of the CPMG modulation function with $\tau=4 \ \mu$s, comparing the instantaneous pulses with no flip angle error (``Ideal'') to finite pulses without errors (``Finite'') and with (``Error'') flip angle errors. See Supplementary information sections S6A and S6C for details of the different sequence conditions. (a) and (b) show a zoomed in view of the first pulse. (c) and (d) show the entire modulation functions $f_x$ and $f_z$ for a 32-pulse sequence with detection at $t=128$ $\mu$s. The black lines in the background highlight the approximately sinusoidal and cosinusoidal modulations to $f_x(t)$ and $f_z(t)$ caused by the flip angle errors. (e) and (f) show a zoomed in view of $f_x(t)$ and $f_z(t)$ for the first 6 pulses to highlight the cumulating error effect of the flip angle error. (g) shows the overlap integral of Equation 7 as a cumulative integral for the ideal and various non-ideal pulses. Here we used a model power spectrum of $S(\omega)= 5\times10^{10} (\textrm{Hz})^2 \omega^{-1}$, and a control sequence with $\tau=4 \ \mu$s. The ``$\delta$ Peak'' data points are the cumulative sum from computing the $\chi$ in the discrete form with Equation 11. (h) shows filter functions calculated numerically from the ideal and non-ideal modulation functions. The $2/(\pi \omega^2 \tau^2)$ dots are the filter peaks in the ideal limit. The inset shows the $S(\omega_1)$ peak in detail, revealing the asymmetric peak splitting caused by the flip angle errors.
}
\end{figure*}

Finite pulses cause the peaks of the filter function in the high-frequency limit to be smaller than that given by Equation 10. To determine the correction, we numerically compute the approximate filter based on the profile of our pulse. The realistic modulation function can be represented as the ideal modulation function convolved with some function $\vec{K}(t)$,
\begin{equation}
    f_\mu(t) = \int_{-\infty}^{\infty}dt' f_\textrm{Ideal}(t) K_\mu(t-t') \ ,
\end{equation}
where $f_\textrm{Ideal}(t)$ is the ideal modulation function and $K_\mu(t)$ is the convolving function for the $\mu$ axis. We do not need to know the exact form of $K_\mu(t)$, since we will be computing $f_\mu(t)$ directly from $\Phi(t)$. However, we note that $K_z(t)$ is roughly Gaussian shaped, and $K_x(t)$ is roughly the derivative of a Gaussian, both with width comparable to the pulse duration. We can use the convolution theorem to write
\begin{equation}
    \mathcal{F}\{f_\mu\}(\omega) = \mathcal{F}\{f_\textrm{Ideal}\}(\omega) \mathcal{F}\{K_\mu\}(\omega),
\end{equation}
where $\mathcal{F}\{f\}$ denotes the Fourier transform of $f$.
Taking the magnitude squared and summing over the components $\mu$ gives us the realistic filter function.
\begin{eqnarray}
    F(\omega) &= \sum_\mu F_\mu(\omega) \\
    &=\sum_\mu \frac{1}{t^2}\left| \mathcal{F}\{f_\mu\}(\omega) \right|^2 \\
    &= \sum_\mu F_\textrm{Ideal}(\omega) A_\mu(\omega) \\
    &= F_\textrm{Ideal}(\omega)A_\textrm{Finite}(\omega)
\end{eqnarray}
where we have defined $A_\mu(\omega) \equiv |\mathcal{F}\{K_\mu\}(\omega)|^2$ and $A_\textrm{Finite}(\omega)\equiv \sum_\mu A_\mu (\omega)$. Figure \ref{fig:FFinite} shows examples of modulation functions and the resulting filter functions with finite pulses.

To obtain $A_\textrm{Finite}(\omega)$, we compute $F(\omega)$ starting from defining $\Phi(t)$ and using Equations \ref{Eqn:ModFuncFromPhi} and \ref{Eqn:FilterMultiAx}, then divide by $F_\textrm{Ideal}(\omega).$ Importantly, $A_\textrm{Finite}(\omega)$ is independent of the interpulse delay $\tau$, so once we have found it for one sequence (CPMG-1 for simplicity), the finite pulse filter $A(\omega)$ can be applied in the analysis of all sequences without the need to reconstruct the modulation function each time.

In the harmonic analysis, we include $A_\textrm{Finite}(\omega)$ as a correction to Equation 21 as follows.
\begin{eqnarray}
\fl \chi(t, \tau) &= \frac{t^2}{2} \int_{-\infty}^{\infty} d\omega F_\textrm{Ideal}(\omega)A_\textrm{Finite}(\omega) \\
\fl &= \frac{4t}{\pi^2} \Biggl[\left( \sum_{m=1,3,5...}^\infty \frac{S_B(m\pi/\tau)A_\textrm{Finite}(m\pi/\tau)}{m^2} \right) + \frac{S_P(\omega_s)A_\textrm{Finite}(\omega_s)}{m_s^2}\Biggr] \\
\fl    &= \chi_B + \frac{4t}{\pi^2}\frac{S_P(\omega_s)A_\textrm{Finite}(\omega_s)}{m_s^2}.
\end{eqnarray}
The background term $\chi_B$ is acquired by the curve fitting described in Section IV.B and therefore needs no explicit correction. We thereby obtain the corrected version of Equation 21.
\begin{equation}
    S_P(\omega_s)
    = \frac{\pi^2 m_s^2}{4 N \tau A_\textrm{Finite}(\omega_s)} 
    \left[\chi(t,\tau) - \chi_B(t,\tau)\right]
\end{equation}

\subsection{Numerical reconstruction of spectra}
\label{sisec:Recon}

We obtain the realistic filter function by construction of $\Phi(t)$ as follows. First, we convolve the programmed pulse waveform profile with the resonator impulse response to approximate the $B_1(t)$ pulse envelope. The impulse response we approximate as $e^{-t|\omega_0|/(2Q)}$, where $|\omega_0| = 2.5 \times 2\pi$ rad GHz is the resonant frequency and $Q$ is the resonator quality factor. In the NV experiments on Sample A, the programmed pulse waveform was a 64 ns Gaussian with $\sigma = 13$ ns, and the resonator had a $Q = 190$; this pulse corresponds to the modulation and filter functions labeled ``Finite A'' in Figure \ref{fig:FFinite}. In the P1 experiment on Sample B, the programmed pulse waveform was a square pulse with length 210 ns, and the resonator had a $Q$ of 2000 (for more sensitive detection of the weak echo); this pulse corresponds to the modulation and filter functions labeled ``Finite B'' in Figure \ref{fig:FFinite}. Since $\omega_1(t) \propto B_1(t)$, and we know the calibrated delays and pulse amplitudes for implementing a $\pi$ pulse, we normalize and center the pulse such that $\Phi(0)=0$, $\Phi(\tau/2) = \pi/2$ and $\Phi(t_f)= \pi$, where time $\tau/2$ is the middle of the pulse, and $t_f$ is some time well after the ring-down of the pulse. Applying Equation \ref{Eqn:ModFuncFromPhi} then produces the modulation function. 

The $x$ and $z$-components of modulation functions with finite pulses are shown in Figure \ref{fig:FFinite}(a) through (f). The lines labeled Finite A and Finite B are for exact $\pi$ pulses with conditions corresponding to the NV and the Sample B P1 experiments respectively. Figure \ref{fig:FFinite}(h) compares the realistic filters against the ideal (``Ideal") filter. At low frequencies $\omega < 1/t_p$, the finite filters and the ideal filter match. At high frequencies $\omega > 1/t_p$, the finite filters decay rapidly. The smooth pulse transition -- analogous to sending a square wave signal through a low-pass filter -- makes the sequence insensitive to high-frequency noise. However, at frequencies near $1/t_p$, the finite filter is slightly greater than the ideal filter by as much as 10\%, peaking in Figure \ref{fig:FFinite}(h) for the Finite A filter around 5 MHz and for the Finite B filter around 0.7 MHz.

In our numerical reconstruction of $S(\omega_1)$ of Figures 4(a), (b) and (c), we generate an array of $\chi$ values for different $\tau$ and detection time $t$ values, similar to the experimental data presented in Figure 2(a), (b), or (c). The values of $\chi$ are calculated with Equation \ref{Eqn:ChiMultAxFilt} using the filter construction method described above and using the power spectrum properties obtained from the power law and \car peak fitting procedures described in Sections IV.B and IV.C. We then perform the contour tracing and power law fit analysis on the numerically generated $\chi$ array in the same manner as performed on an experimentally obtained $\chi$ array.

\subsection{Modeling}
\label{sisec:FiniteErrors}

It is convenient to use the $\delta$-function peak approximation of the filter (Equation 10) in CPMG analyses, however, in the presence of finite pulses or flip angle errors, there can be significant deviations between what is observed experimentally and what is predicted by the approximation. Figure \ref{fig:FFinite} shows the modulation functions, filters, and $\chi$ integrals of sequences with finite pulses and flip angle errors. The flip angle error analysis was achieved by altering $\Phi(t)$ such that each pulse increases $\Phi(t)$ by an increment of $\beta \neq \pi$. The modulation and filter functions labeled ``Error A'' are produced from a pulse that has the same duration as ``Finite A'', but has a greater amplitude such that there is an over-rotation of $\beta=1.05\pi$. The functions labeled ``Error B'' likewise correspond to ``Finite B'' but have an under-rotation of $\beta=0.95\pi$. Figure \ref{fig:FFinite}(f) shows how $f_z(t)$ for both Error A and B reach the same level during the interpulse delay, and that this level is incremented with each pulse. On the other hand, Figure \ref{fig:FFinite}(e) shows that the functions $f_x(t)$ during the interpulse delays have opposite sign for the over- and under-rotations. Figures \ref{fig:FFinite}(c) and (d) show $f_x(t)$ and $f_z(t)$ over a longer time interval. With a small error, the modulation functions for Error A and B can be seen producing an additional modulation to the ideal $f_z(t)$ modulation function that is roughly sinusoidal for $f_x(t)$ and cosinusoidal for $f_z(t)$.
\begin{eqnarray}
    f_{x,\textrm{Err}}(t) &\approx f_z(t) \sin\left(\omega_\textrm{mod}t\right) \\
    f_{z,\textrm{Err}}(t) &\approx f_z(t) \cos\left(\omega_\textrm{mod}t\right),
\end{eqnarray}
where $\omega_\textrm{mod}=(\beta-\pi)/\tau$.
In the frequency domain this is equivalent to the ideal filter being convolved with a pair of $\delta$-functions.
\begin{eqnarray}
\fl        F_{x,\textrm{Err}}(t) \approx &\frac{1}{t^2}\bigg| \int_{-\infty}^{\infty} \mathcal{F}\{f_z\}(\omega-\omega') 
        \frac{1}{2i}\left[ \delta\left(\omega - \omega_\textrm{mod}\right) - \delta\left(\omega + \omega_\textrm{mod} \right) \right] d\omega' \bigg|^2 \\
\fl        F_{z,\textrm{Err}}(t) \approx &\frac{1}{t^2}\bigg| \int_{-\infty}^{\infty} \mathcal{F}\{f_z\}(\omega-\omega') \frac{1}{2}\left[ \delta\left(\omega - \omega_\textrm{mod}\right) + \delta\left(\omega + \omega_\textrm{mod} \right) \right] d\omega' \bigg|^2
\end{eqnarray}
Further distortions arise due to the asymmetry of the pulse, and due to the modulation of the $x$-component of the modulation function being a sine modulation at the modulation frequency $\omega_\textrm{mod}$. The inset of Figure \ref{fig:FFinite}(h) shows how the flip angle error results in a splitting of the peaks in the filter function. This feature appears as a general result of the control, without any knowledge of the system-environment Hamiltonian. It has been shown recently with Floquet analysis that over/under-rotations can be used for polarization and readout of a single \car via a single NV center \cite{lang_quantum_2019, whaites_adiabatic_2022}.

\begin{figure}[ht]
\includegraphics[width=0.6\linewidth]{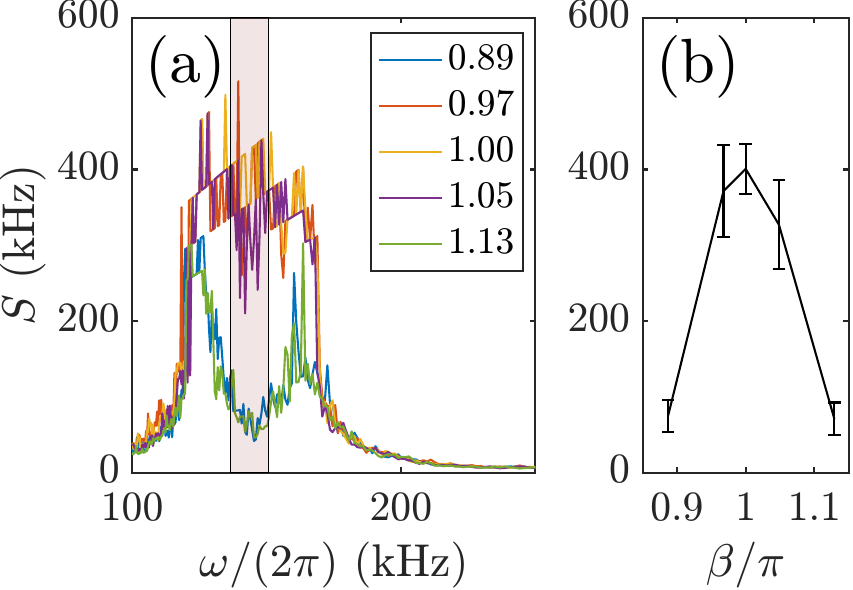}
\caption{\label{fig:F_PulseAmpCheck13} (a) The $S(\omega_1)$ spectra of the \car peak at 13 mT obtained from NV experiments where the cycle pulse amplitudes were varied. The numbers in the legend are $\beta/\pi$, where $\beta$ is the rotation angle of the CPMG cycle pulse. (b) The mean (plotted value) and standard deviation (error bars) of each peak -- sampled in the shaded region of (a) -- are shown as a function of $\beta/\pi$.}
\end{figure}

\begin{figure*}[ht]
\includegraphics[width=\textwidth]{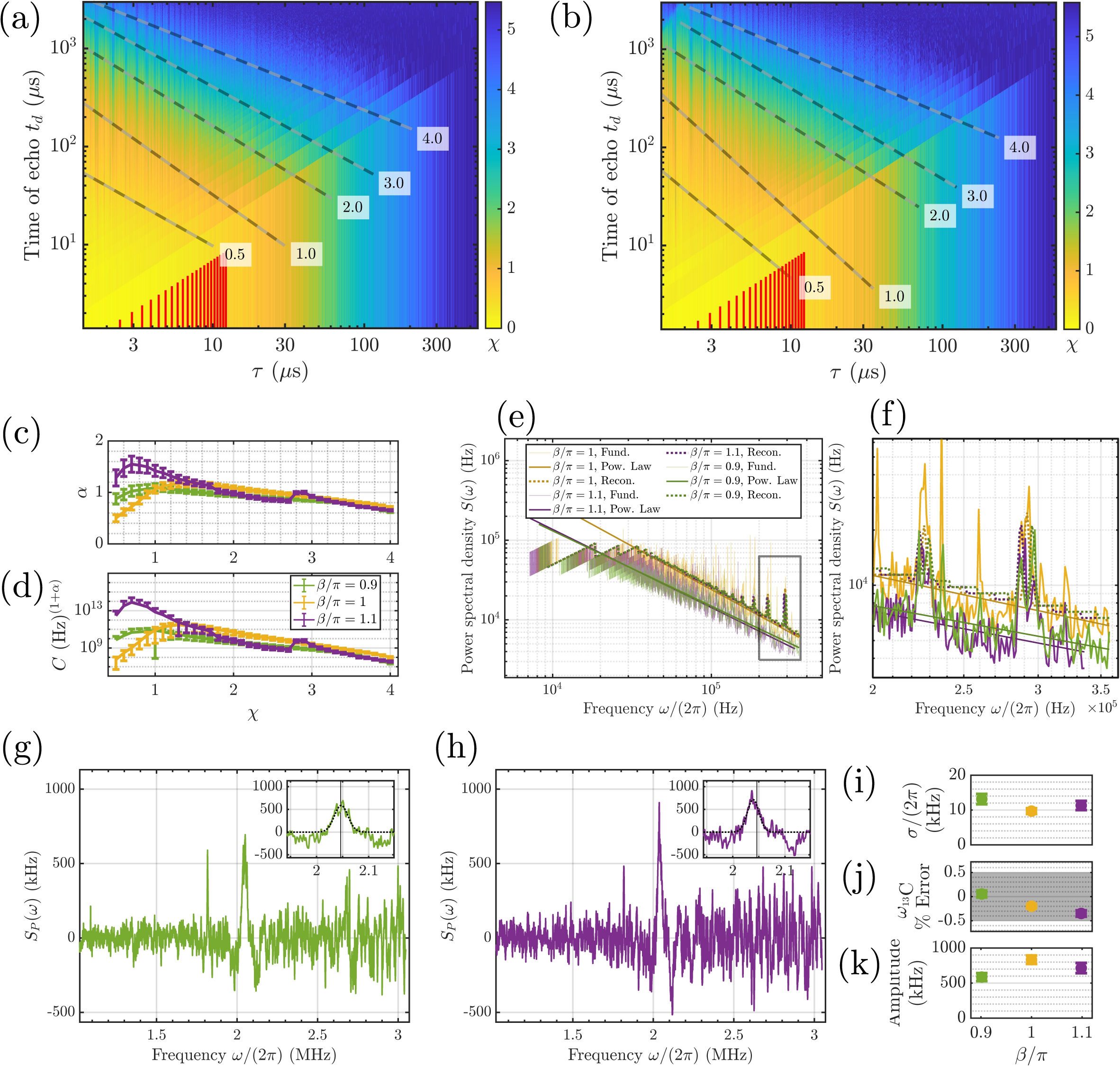}
\caption{\label{fig:F2_190Comp} Analyses similar to those presented in Figures 2 -- 5 of the main text are applied to CPMG experiments on NV centers at 190 mT with 10\% over- and under-rotations. Image plots of $\chi$  are shown for the 10\% under-rotation (a) and the 10\% over-rotation (b). Black and white dashed lines represent the power law fits, and the red vertical stripes indicate revival $\tau$ values as in Figure 2. (c) and (d) compare the $\alpha$ and $C$ power law ($S(\omega)=C\omega^{-\alpha}$) fit values for the under- and over-rotation experiments ($\beta/\pi = 0.9$ and $\beta/\pi = 1.1$ respectively) to the experiment with no flip angle error ($\beta/\pi = 1$). (e) shows the $S(\omega_1)$ for the $\chi=3.5$ contour, the corresponding fit power law spectra (``Pow. Law''), and the numerically reconstructed spectra (``Recon.''). (f) shows the boxed portion of (e) in greater detail, revealing the asymmetric peak-splitting effects caused by under- and over-rotations. (g) and (h) show the spectra of the \car peak obtained with the harmonic analysis procedure for the under- and over-rotation experiments respectively. (i), (j), and (k) compare the harmonic analysis characterization of the \car peak width, frequency percentage error, and amplitude.}
\end{figure*}

\begin{figure}[ht]
\includegraphics[width=0.5\linewidth]{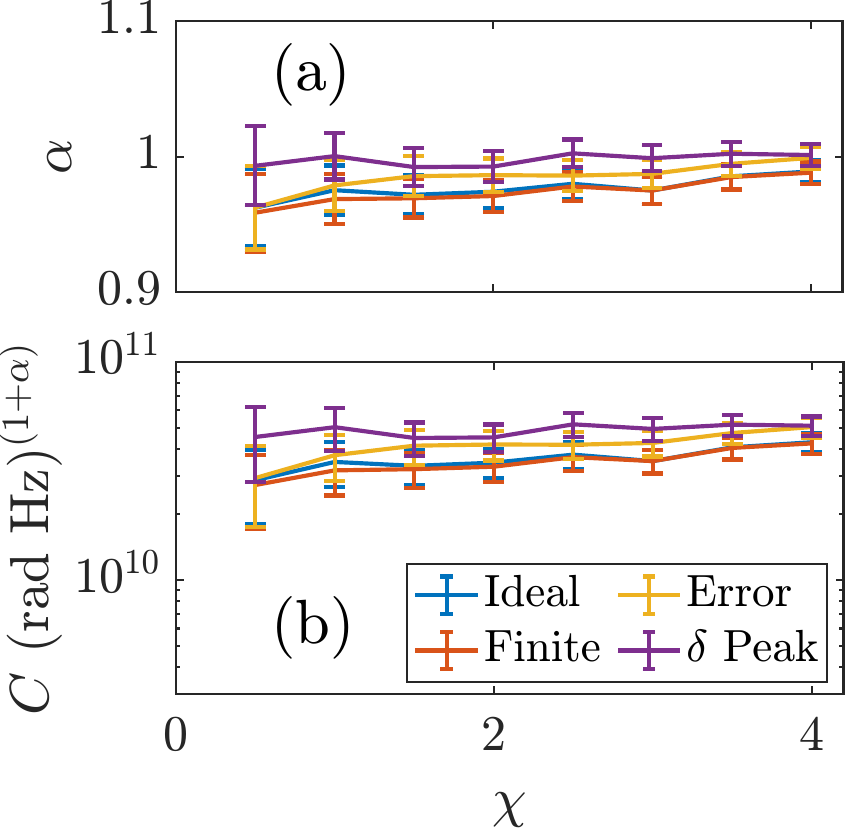}
\caption{\label{fig:FFiniteSimComp} Results from numerically computing $\chi$ arrays from a model power law spectrum $S(\omega)= 5\times10^{10} (\textrm{Hz})^2 \omega^{-1}$ and ideal and non-ideal filter functions. $\alpha$ in (a) and $C$ in (b) are the recovered fit values from analyzing the arrays with the same procedure that was performed for the real experimental data.
}
\end{figure}

We also performed a numerical calculation to assess whether the flip angle errors affect the power law spectrum measurement. Using a model spectrum of $S(\omega) = C \omega^{-\alpha}$ with $C=5\times10^{10} \, (\textrm{Hz})^{(1+\alpha)}$ and $\alpha = 1$, and filter functions calculated in the method described above, we generated $\chi$ arrays and performed the power law fit analysis (Equation 10). The results are shown in Figure \ref{fig:FFiniteSimComp}. The four lines correspond to different conditions for generating the filter function. For the line labeled ``Ideal", the modulation function consisted of sharp transitions from instantaneous pulses. For the line labeled ``Finite'', the pulse duration was the same as conditions for Finite A, listed above. For the line labeled ``Error'', the modulation function was generated with the same finite width as for ``Finite'', but with an over-rotation of 10\%. Lastly, for the line labeled ``$\delta$ Peak'', the filter was generated purely from using the $\delta$-function approximation of Equation 10. The same fitting procedure was applied to the four simulated data sets. The power law fitting equation (Equation 17) is based on the $\delta$-approximation of the filter and does not take into account the finite pulse or flip angle error effects. The fact that the $C$ and $\alpha$ fits all significantly overlap suggests that the finite pulse widths and flip angle errors do not drastically affect the power law fitting results. However, the numerical construction of the filters in this investigation was carried out only to the leading order term of the Magnus expansion, and it is uncertain if incorporating the higher-order terms would lead to significant deviations.

\subsection{Experiments}

We have examined the peak-splitting effect due to flip angle errors experimentally by performing CPMG on NV centers at 13 mT and varying the amplitude of the cycle pulse. We varied the value of $\tau$ to sweep the fundamental modulation frequency over the \car peak. Results of the $S(\omega_1)$ with $\chi=3$ are shown in Figure \ref{fig:F_PulseAmpCheck13}. We characterize the \car peak amplitude as the mean of the points within $\pm5\%$ of the \car Larmor frequency. The results suggest that for flip angle errors of $\lesssim5\%$, the peak is only slightly diminished. However, this might be partially obscured by the fact that at the \car peak, the data points in the spectrum are obtained at low echo number $N$. Nevertheless, we see that for flip angle errors of $\gtrsim10\%$, the peak is strongly split, with the inversion in the middle being reduced to as little as 20\% of the full peak amplitude.

To explore how the power law characterization and the \car peak harmonic analysis are affected by flip angle errors, we have performed two experiments at 190 mT with over and under rotations of 10\% achieved by adjusting the amplitude of the pulse. Results are shown in Figure \ref{fig:F2_190Comp}. We have performed the same analyses that we have performed on the experiments presented in the main text with no flip angle error. Recall that this analysis assumes the $\delta$-function approximation of the filter in fitting the power law, and it includes the finite pulse correction factor $A_\textrm{Finite}(\omega)$ in the harmonic analysis but no flip angle correction. Figures \ref{fig:F2_190Comp}(c) and (d) show the power law fit parameters. At low $\chi\lesssim1$, the sequences with flip angle errors appear to measure a spectrum with larger $C$ and $\alpha$. However, for the majority of $\chi$ values, there is mostly overlap between the fits, suggesting no significant deviation of the $C$ and $\alpha$ measurement in the presence of small flip angle errors. For the \car peak characterization, the differences in fit parameters are slightly more significant. Consistent with the realistic filter simulation -- which is seen in the inset of Figure \ref{fig:FFinite}(h) to widen the filter peak -- the measurement with the $\beta \neq \pi$ sequences indicates greater width and lower amplitude of the \car peak than does the measurement with no flip angle error. The asymmetric splitting of the peak appears to lead to a shift in frequency estimation of the \car peak. However, the deviation in the frequency estimate obtained from the harmonic analysis is within the uncertainty of the frequency value based on experimental conditions as discussed in Section IV.C.

\section{Estimating $(S(\omega \rightarrow 0))$}
\label{sisec:ZeroFreqLimit}
We can estimate the finite magnitude of the power spectrum in the limit $\omega \rightarrow 0$ by measurements of $T_2$ obtained from Hahn echo experiments \cite{yuge_measurement_2011}. 
In the limit $\tau \rightarrow \infty$, the peaks of the filter function approach $\omega \rightarrow 0$. In this limit, Equation 11 leads to $S(0) = 2 \chi(t)/t$, which differs only slightly from Equation 12 by a factor of $\pi^2/8 \approx 1.2$. The Hahn echo $T_2$ values are provided in Section \ref{sisec:Samples}. The resulting $S(0)$ are 41(11) kHz for the NV at 13 mT, 82(1) kHz for the NV at 190 mT, 2.4(2) MHz for the Sample A P1 center, and 11(2) kHz for the Sample B P1 at 89 mT. These values are in good agreement with the low-frequency levels of the respective spectra in Figure 4.

\newpage

\section*{References}
\bibliographystyle{unsrt}
\bibliography{supplementary}